\documentclass[aps,prb,twocolumn,superscriptaddress,showpacs,a4paper]{revtex4}
\usepackage{amsfonts,amsmath}
\usepackage[T1]{fontenc}
\usepackage[utf8]{inputenc}
\usepackage{color}
\usepackage{ifthen}
\usepackage{ulem}

\usepackage[dvips]{graphicx}
\usepackage[hypertex,breaklinks=true,colorlinks=false]{hyperref}
\newcommand{\ext}{eps}

\newcommand{\journal}[4]
{\ifthenelse{\equal{#1}{prl}}{
\prl {\bf #2}, \href{http://link.aps.org/abstract/PRL/v#2/e#3}{#3} (#4)}
{\ifthenelse{\equal{#1}{prb}}{
\prb {\bf #2}, \href{http://link.aps.org/abstract/PRB/v#2/e#3}{#3} (#4)}
{\ifthenelse{\equal{#1}{arxiv}}{preprint
\href{http://arxiv.org/abs/#2.#3}{arXiv:#2.#3}}
{\ifthenelse{\equal{#1}{rmp}}{
\rmp {\bf #2}, \href{http://link.aps.org/abstract/RMP/v#2/e#3}{#3} (#4)}
{\ifthenelse{\equal{#1}{cond-mat}}{preprint
\href{http://arxiv.org/abs/cond-mat/#2}{cond-mat/#2}}
{\ifthenelse{\equal{#1}{pre}}{
\pre {\bf #2}, \href{http://link.aps.org/abstract/PRE/v#2/e#3}{#3} (#4)}
{#1 {\bf #2}, #3 (#4)}}}}}}}

\newcommand{\npb}{Nucl. Phys. B}

\newcommand{\journaldoi}[5]{#1\ {\bf #2}, \href{http://dx.doi.org/#5}{#3} (#4)}

\newcommand{\la}{\langle}
\newcommand{\ra}{\rangle}

\definecolor{ca}{rgb}{1,0,0}    

\definecolor{cc}{rgb}{0.1,0.1,.6}    

\begin{document}

\title{
Shannon and entanglement entropies of one- and two-dimensional critical wave
functions
}
\date{05/10/2009}
\author{Jean-Marie St\'ephan}
\affiliation{
Institut de Physique Th\'eorique,
CEA, IPhT, CNRS, URA 2306, F-91191 Gif-sur-Yvette, France.}
\author{Shunsuke Furukawa}
\affiliation{
Condensed Matter Theory Laboratory, RIKEN, Wako, Saitama 351-0198, Japan
}
\author{Gr\'egoire Misguich}
\affiliation{
Institut de Physique Th\'eorique,
CEA, IPhT, CNRS, URA 2306, F-91191 Gif-sur-Yvette, France.}
\author{Vincent Pasquier}
\affiliation{
Institut de Physique Th\'eorique,
CEA, IPhT, CNRS, URA 2306, F-91191 Gif-sur-Yvette, France.}

\begin{abstract}

We study 
the Shannon entropy of the probability distribution  
resulting from the ground-state wave function of a one-dimensional quantum
model.
This entropy is related to the entanglement entropy of a Rokhsar-Kivelson-type
wave function  
built from the corresponding two-dimensional classical model.
In both critical and massive cases, 
we observe that it is composed of an extensive part proportional to the length
of the system 
and a subleading universal constant $S_0$.
In $c=1$ critical systems (Tomonaga-Luttinger liquids), 
we find that $S_0$ is a simple function of the boson compactification radius. 
This finding is based on a field-theoretical analysis 
of the Dyson-Gaudin gas 
related to dimer and Calogero-Sutherland models. We also performed numerical demonstrations 
in the dimer models and the spin-$1/2$ XXZ chain.
In a massive (crystal) phase, $S_0$ is related to the ground-state degeneracy.
We also examine this entropy in the Ising chain in a transverse field as an
example showing a $c=1/2$ critical point.

\end{abstract}

\pacs{05.50.+q, 75.10.Pq, 03.67.Mn}


\maketitle

\section{Introduction}
\label{sec:intro}

There has been growing interest in quantifying entanglement in extended quantum
systems 
to detect non-trivial correlations existing in many-body ground
states.\cite{afov08}
A useful measure of entanglement is the von Neumann entanglement entropy 
$S^{\rm VN}(A) := -{\rm Tr}~\rho_A \log \rho_A$, 
defined from the reduced density matrix $\rho_A$ of a subsystem $A$.  
Its novelty lies in its universal behavior reflecting the long-distance nature of
the system. 
In one-dimensional critical systems, for instance, 
the entanglement entropy of a long interval of length $\ell$ 
shows a universal scaling\cite{hlw94,vlrk03,korepin04,cc04} 
$S^{\rm VN} (A)\simeq \frac {c}{3} \log \ell + {\rm const.}$,
where $c$ is the central charge of the conformal field theory (CFT) 
describing the long distance correlations. 
Possible further information of CFT can be encoded in a multi-interval
entanglement entropy.\cite{ch04,fps09,cct09}
As another example, the existence of topological order in gapped systems can be
detected 
by measuring a constant contribution to the entanglement
entropy\cite{kp06,lw06} 
(with recent applications to fractional quantum Hall states \cite{hzs07,fl08}
and $\mathbb{Z}_2$ spin liquids \cite{fm07,hiz05,cc07,prf07,hzhl08}).

Here we introduce an {\it apparently} different entropy as follows.
Consider a one-dimensional (1D) quantum model and its ground state $|g\rangle$.
If one chooses an orthogonal basis $\{|i\rangle\}$ of the Hilbert space, 
one gets a set of probabilities $p_i:=|\la i|g\ra|^2$,
from which a Shannon entropy can be defined:
\begin{equation}
 S:=-\sum_i p_i \log p_i
 \label{eq:Sdef}.
\end{equation}
Note that this entropy depends on the choice of basis. 
For a $U(1)$-symmetric model with the conservation of the particle number or the
magnetization, 
we use the local particle occupations $\{n_j\}$ or magnetizations
$\{\sigma_j^z\}$ to define the basis. 
The entropy $S$ is small when the wave function $|g\ra$ is dominated by a
particular crystal state $|i_0\ra$.
It becomes larger 
as more basis states contribute to the wave function due to quantum
fluctuations. 
Thus, this entropy quantifies quantum fluctuations or entanglement occurring in
the given basis.
Like other entanglement measures, we will see that 
the scaling of this entropy is controlled by the essential long-distance nature
of the system.
We note that a similar entropy also appears 
in the context of dynamical systems,\cite{Anantharaman,Sinai} where it is
used to quantify chaos.
\footnote{A measure of phase space which is invariant under a chaotic transformation can sometimes be
represented by a probability distribution on the configurations of a spin chain.}

At the same time, 
this entropy has an interpretation 
as the entanglement entropy of a two-dimensional (2D) quantum state,  
as we describe in detail in Sec.~\ref{sec:RK_states}. 
The basic idea goes as follows.  
A 1D quantum Hamiltonian on a ring is related to a 2D classical model on a
cylinder 
in the transfer matrix formalism.
Then a Rokhsar-Kivelson (RK) -type wave function\cite{rk88,henley04} $|{\rm
RK}\ra$
can be constructed from this 2D classical model.
One can show that in the limit of a long cylinder, the entropy $S$ defined in
Eq.~\eqref{eq:Sdef} is precisely 
the entanglement entropy $S^{\rm VN}(A)$ of the 2D RK state $|{\rm RK}\ra$
for a half cylinder $A$ shown in Fig.~\ref{fig:AB}.
More specifically, 
each probability $p_i$ becomes an eigenvalue of the reduced density matrix
$\rho_A$. 
This correspondence allows us to apply a simpler 1D picture 
to study the universal behavior of entanglement entropy in 2D critical systems, 
a very active subject in recent literature.\cite{fm06,hsu08,mfs09}  

The purpose of the present paper is to unveil the generic scaling properties 
of the Shannon entropy \eqref{eq:Sdef} of 1D ground states $|g\ra$, 
as a function of the ring length $L$.  
This amounts to studying the entanglement entropy $S^{\rm VN}$ 
of 2D RK states $|{\rm RK}\ra$ defined on a cylinder of circumference $L$. 
In both critical and massive systems, 
we observe that for large $L$, the entropy is composed of 
an extensive part proportional to $L$
and a subleading constant:
\begin{equation}
 S(L) = \alpha L + S_0 + o(1).
\end{equation}
The extensive part $\alpha L$ simply reflects the fact 
that a generic wave function $|g\rangle$ spreads 
over an exponentially large number of microscopic configurations.
In terms of the 2D entanglement entropy, this can be interpreted as a boundary
contribution. 
Here, we are however interested in the subleading constant $S_0$. 
As we will see, $S_0$ is universal and is determined by the basic properties of
critical or massive systems.

Our primary interest lies in the situation 
where a 1D quantum or 2D classical system (used to build a RK state) is 
described by a $c=1$ massless bosonic field theory (Tomonaga-Luttinger liquid; TLL)
with the Euclidean action:
\begin{equation}
\mathcal{A[\phi]}=\frac{1}{8\pi}\int\int dx dy \left[ (\partial_x
\phi)^2+(\partial_y \phi)^2 \right].
\label{eq:FreeFieldAction}
\end{equation}
Here the bosonic field is compactified on a circle: $\phi\equiv\phi+2\pi R$. 
The boson compactification radius $R$ is an important scale-invariant number 
which controls the power-law behavior of various physical quantities.
\footnote{
Here we take a standard notation used in statistical mechanics and field
theory, 
where $R=1$ for a free fermion and $R=\sqrt{2}$ for a $SU(2)$-symmetric case. 
In condensed matter physics, a common notation is 
$R^{\rm cm}=R/\sqrt{4\pi}$, 
and the coupling constant $K=1/R^2$ known as TLL parameter is also widely used. 
} 
We find that $S_0$ is given by a simple function of the radius:
\begin{equation}
 S_0=\log R -\frac{1}{2}.
 \label{eq:S0}
\end{equation}
We present detailed analyses to establish this result. In Sec.~\ref{sec:RKtoGaudin}, 
we study the Dyson-Gaudin Coulomb gas model,\cite{dyson,gaudin}
which gives probabilities $\{p_i\}$ 
for the dimer model on the hexagonal lattice
and the Calogero-Sutherland model.\cite{Suth1,Suth2} 
In particular, in Sec. \ref{sec:field}, 
we analytically derive Eq.~\eqref{eq:S0} 
using a free field representation of the gas model. 
In Sec.~\ref{sec:xxz_chain}, we numerically demonstrate 
the same result in the spin-$1/2$ XXZ chain.

At a certain value of $R$, the system undergoes a phase transition 
to a massive crystal phase. 
As we show in Sec. \ref{sec:transition} and Sec. \ref{sec:xxz_chain}, 
in the massive phase breaking a symmetry, 
$S_0$ is related to the ground-state degeneracy~$d$:
\begin{equation}
 S_0 = \log d.
\end{equation}
At the transition point, we observe a jump in $S_0$, 
though it is slightly obscured due to finite-size effects.

As an example showing the $c=1/2$ criticality, in Sec.~\ref{sec:ising_chain},
we study an Ising chain in a transverse field~\eqref{eq:Hictf}. 
We calculate the entropies in $\sigma^z$ and $\sigma^x$ bases, 
corresponding to the RK states built from an eight-vertex model and a 2D Ising
model, respectively.   
The extracted constant $S_0^{(z)}\simeq -0.4387$ and $S_0^{(x)}\simeq 0.2544$
at the critical point 
might be generic constants characterizing the $c=1/2$ CFT, 
although we do not have any analytical derivation of these numbers. 
In the symmetry-broken phase, the constant takes $S_0^{(z)} = -\log 2$ in the
$\sigma^z$ basis, 
which is interpreted as a manifestation of $\mathbb{Z}_2$ topological order 
in the eight-vertex RK state.

As a related quantity, 
in Sec.~\ref{sec:scaling_pmax}, 
we study the scaling of the largest probability $p_0:=\max p_i$.
This maximum is attained by crystal states $|i_0\ra$, 
e.g., by N\'eel states 
$|\!\!\uparrow\downarrow\uparrow\downarrow\!\dots\ra$ and 
$|\!\!\downarrow\uparrow\downarrow\uparrow\!\dots\ra$ 
for the XXZ chain in zero magnetic field. 
Very similarly to the entropy $S(L)$, 
the logarithm of $p_0$ has 
a dominant linear contribution followed by a subleading constant: 
\begin{equation}
 -\log p_0=\tilde{\alpha} L + \gamma+o(1). 
\end{equation}
For $c=1$ critical systems, our numerical results 
in the Dyson-Gaudin gas and the XXZ chain 
show that
\begin{equation}\label{eq:gamma}
 \gamma=\log R.
\end{equation}
This result, together with Eq.~\eqref{eq:S0}, 
gives a simple and universal way to determine the radius $R$ 
from a ground state wave function.

Here we comment on closely related works. 
Using boundary CFT, 
Hsu {\it et al.}\cite{hsu08} have also studied 
the entanglement entropy of 2D RK wave functions for a half cylinder. 
Their prediction for the constant $S_0$ differs from ours, although it matches 
our calculation for a different constant $\gamma$ appearing in $-\log p_0$.
A quantity similar to $p_0$ has also been studied by Campos Venuti {\it et
al.}\cite{cvsz09} in the context of fidelity, 
in agreement with our result \eqref{eq:gamma}.

\section{Correspondence between Shannon and entanglement entropies}
\label{sec:RK_states}

In this section, we formulate the connection 
between the entanglement entropy of Rokhsar-Kivelson states 
and the Shannon entropy of the ground states of 1D quantum models.

\subsection{Generalized Rokhsar-Kivelson states}

We start from a discrete classical (spin) model on a 2D lattice 
defined by Boltzmann weights $e^{-E(c)}$ for microscopic configurations $c$ of
the system. 
The partition function is given by  
\begin{equation}
 \mathcal{Z}=\sum_c e^{-E(c)}.
 \label{eq:Zcl}
\end{equation}
The Hilbert space of a 2D quantum system is constructed by associating a basis
state $|c\ra$
to each classical configuration $c$.
Then, one can define a generalized Rokhsar-Kivelson (RK) wave function as 
the linear combination of all the basis states $|c\ra$ 
with amplitudes given by 
the square roots of the classical Boltzmann weights: \cite{rk88,henley04}
\begin{equation}
 | {\rm RK} \rangle = \frac{1}{\sqrt \mathcal{Z}} \sum_c e^{-\frac{1}{2} E(c)}
|c\rangle.
	\label{eq:RK}
\end{equation}
The RK state shares the same correlations with the original classical model, 
as long as one focuses on the diagonal correlations in the $\{ |c\ra \}$-basis. 
This type of state was first introduced as the ground state of the quantum dimer
model (QDM) on the square lattice, 
where $\{c\}$ are fully packed dimer coverings of the lattice. 
Later, the same type of states have been studied 
for different lattices (hexagonal,\cite{msc01} triangular,\cite{ms01},
kagome\cite{msp02}, etc.) 
and for a modified dimer model.\cite{ccmp07}
The RK states for the eight-vertex model on the square lattice 
have also been studied.\cite{aff04}

\begin{figure}
\begin{center}
\includegraphics[height=4cm]{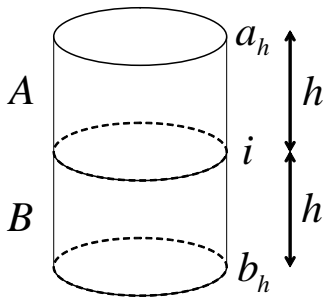}
\caption[1]{Two dimensional system with a cylinder geometry, divided into upper
and lower parts, $A$ and $B$.
}\label{fig:AB}
\end{center}
\end{figure}

\subsection{Schmidt decomposition and entanglement entropy}

We divide the system into two parts, $A$ and $B$, as in Fig.~\ref{fig:AB}.
The reduced density matrix $\rho_A$ is obtained
from a state $|{\rm RK}\rangle$ by tracing out the degrees of freedom in  $B$:
\begin{equation}
 \rho_A = \mathop{\rm Tr}_{B} ~| {\rm RK} \rangle \langle {\rm RK} |.
\end{equation}
We are interested in the (Von Neumann) entanglement entropy of $A$:
\begin{equation}
 S^{\rm VN}(A)=-{\rm Tr} ~\rho_A \log \rho_A.
\end{equation}
Here we show that 
the calculation of $S^{\rm VN} (A)$ for a RK state \eqref{eq:RK} 
can be recast as a fully classical calculation, 
provided that the boundary between $A$ and $B$ satisfies certain geometrical
conditions.
This is done by deriving a Schmidt decomposition of the RK state \eqref{eq:RK}.

\subsubsection{Case with local constraints}
\label{sec:schmidt1}
\begin{figure}
\begin{center}
\includegraphics[height=5cm]{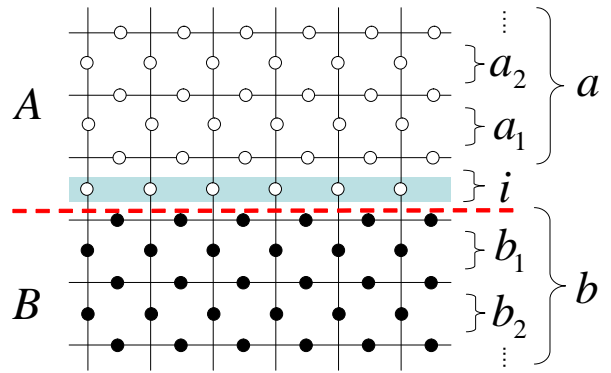}
\caption[1]{(color online) 
Spatial division of the square lattice into regions $A$ and $B$, 
corresponding to the procedure in Sec.~\ref{sec:schmidt1}. 
The system is defined on a cylinder, and 
is periodic in the horizontal direction. 
}\label{fig:AB_aib}
\end{center}
\end{figure}

We first consider the case where the classical model contains 
certain local constraints.
For simplicity, we assume that the system consists of Ising variables $\sigma_j$
sitting on the bonds of the square lattice as in Fig.~\ref{fig:AB_aib} 
(The same argument also applies to a model on the hexagonal lattice.). 
Around each site, the four Ising variables $\sigma_j$'s satisfy the following
constraint:
if three of them are specified, the last one is uniquely determined. 
Dimer, six-vertex, and loop models satisfy this condition.
For example, in the case of a dimer model, 
we assign $\sigma_j=\pm 1$ to the presence/absence of a dimer on the bond $j$. 
Then, there is strictly one bond with $\sigma_j=+1$ emanating from each site.

We define the system on a cylinder, 
and divide it into $A$ and $B$, 
as shown in Fig.~\ref{fig:AB_aib}.
Here, all spins at the boundary belong to $A$. 
The spin configurations $a$ and $b$ inside $A$ and $B$ must agree 
with the configuration $i$ at the boundary. 
Let ${\cal E}_i^A$~(${\cal E}_i^B$) be a set of such $a$'s ($b$'s).
Thanks to the local constraints, 
such sets share no common element:
\begin{equation}\label{eq:set_i_omega}
 {\cal E}_i^\Omega \cap {\cal E}_{i^\prime}^\Omega = \emptyset~~
 (\Omega=A,B;~ i\ne i^\prime).
\end{equation}

We assume that the classical model $E(c)$ contains only 
local interactions involving four bonds around each site.
The energy can then be decomposed into two parts: 
\begin{equation}
 E(c)=E_A(a,i) +E_B(b,i).
\end{equation}
The first term corresponds to all the  interactions among spins in $A$. 
The second corresponds to interactions among spins in $B$ and the boundary
region.
The important point is the absence of direct interaction between spins inside
$A$ and $B$. 
Thanks to this property, the Boltzmann weight of the configuration $c=(a,i,b)$
factorizes into two parts, 
which allows us to rewrite Eq.~\eqref{eq:RK} as follows:
\begin{eqnarray}
	|{\rm RK} \rangle = \frac{1}{\sqrt \mathcal{Z}}
	\sum_i 
    \left[ \sum_{a\in {\cal E}_i^A} e^{-\frac{1}{2} E_A(a,i)}|a,i\ra \right]
\notag \\
	\times
    \left[ \sum_{b\in {\cal E}_i^B} e^{-\frac{1}{2} E_{B}(b,i)}|b\ra \right].
\end{eqnarray}
We define normalized RK states (boundary-dependent) on $A$ and $B$ as 
\begin{subequations}
\label{eq:RK_region}
\begin{eqnarray}
 |{\rm RK}_i^A \ra &:=& \frac1{\sqrt{\mathcal{Z}_i^A}} 
 \sum_{a\in {\cal E}_i^A} e^{-\frac{1}{2} E_A(a,i)}|a,i\ra, 
   \label{eq:RK_A} \\
 |{\rm RK}_i^B \ra &:=& \frac1{\sqrt{\mathcal{Z}_i^B}}
 \sum_{b\in {\cal E}_i^B} e^{-\frac{1}{2} E_{B}(b,i)}|b\ra,
   \label{eq:RK_B}\\
 \text{with} &~&  
 \mathcal{Z}_i^{\Omega}=\sum_{\omega\in \mathcal{E}_i^\Omega}
e^{-E_\Omega(\omega,i)}
 ~~(\Omega=A,B).
\end{eqnarray}
\end{subequations}
Then we arrive at the Schmidt decomposition 
\begin{equation}
	|{\rm RK} \rangle = \sum_i
	\sqrt{p_i}~
	| {\rm RK}_i^A \ra | {\rm RK}^{B}_i \ra,	
	~~\text{with}~~
    p_i := \frac{\mathcal Z_i^{A}\mathcal Z_i^{B}}{\mathcal{Z}}.
	\label{eq:schmidt}
\end{equation}
Here, the mutual orthogonality 
$\langle {\rm RK}_i^\Omega | {\rm RK}_{i^\prime}^\Omega \rangle =
\delta_{ii^\prime}$ 
is guaranteed by Eq.~\eqref{eq:set_i_omega}.
The reduced density matrix $\rho_\Omega $ (with $\Omega=A$ or $B$) is then\footnote{
It is interesting to notice that $\rho_\Omega$ ($\Omega=A,B$) is block-diagonal in the spin-basis,
 each block being labeled by a spin configuration $i$ at the boundary. Block number $i$ is a
 $\textrm{Card }{\cal E}_i^\Omega \times \textrm{Card }{\cal E}_i^\Omega$ matrix, and has only \textit{one} 
non-zero eigenvalue $p_i$, the corresponding eigenvector being $|RK_i^\Omega\ra$.
 In the special case where all the energies $E_\Omega(\omega,i)$ vanish, each
state $|{\rm RK}^\Omega_i\rangle$ is an equal-amplitude superposition of spin configurations.
Then, each block of $\rho_\Omega$ is particularly simple since all its matrix elements are {\it identical}.
The block labeled by the boundary configuration $i$, of size
 $\textrm{Card }{\cal E}_i^\Omega \times \textrm{Card }{\cal E}_i^\Omega$,
 has all its entries equal to $p_i/\textrm{Card }{\cal E}_i^\Omega$
and a trace equal to $p_i$.}
\begin{equation}
    \rho_\Omega=\sum_i p_i | {\rm RK}^\Omega_i \rangle \langle {\rm RK}^\Omega_i |.
\end{equation}
Therefore, we get 
\begin{eqnarray} \label{eq:S_RK}
 S^{\rm VN} (A) =- \sum_i p_i \log p_i.
\end{eqnarray}
The entanglement entropy can thus be computed 
from the classical partition functions ${\cal Z}_i^A {\cal Z}_i^B$ 
with boundary spins fixed in a state $i$.  
Similar formulations were used in Refs.~\onlinecite{hiz05,fm07,cc07,prf07} 
for exact/perturbative/numerical calculations of entanglement entropy 
in toric code, quantum eight-vertex, and quantum dimer models.

\subsubsection{General case}
\label{sec:schmidt2}
\begin{figure}
\begin{center}
\includegraphics[width=8cm]{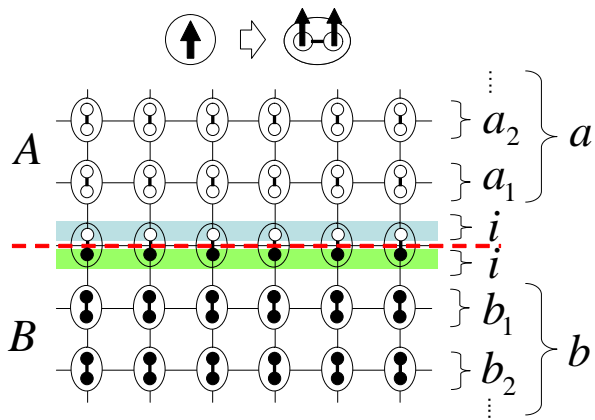}
\caption[1]{(color online) 
Spatial division of the square lattice, 
corresponding to the procedure in Sec.~\ref{sec:schmidt2}. 
Each spin of the original model is replaced by a {\it pair} of spins. 
An infinitely strong ferromagnetic interaction
ensures that the two spins are always in the same state.
The (new) lattice is divided in two regions, $A$ and $B$,
and the original sites which overlap with both regions are called ``boundary''
sites.
The RK state admits a simple Schmidt decomposition (see text) if
the only interactions between $A$ and $B$ take place inside the boundary region.
For general short ranged interactions (not necessarily first neighbor), this
condition can be achieved by choosing a sufficiently ``fattened'' boundary
region.
}\label{fig:AB_aiib}	
\end{center}
\end{figure}

The above discussion relies on the presence of local constraints. 
Without them, any configuration $b$ is allowed in $B$ irrespective of $i$, 
and states $\{ |{\rm RK}_i^B \ra\}$ defined in Eq.~\eqref{eq:RK_B} are not mutually
orthogonal in general.
Even in such a general case, one can slightly modify the model 
so that the entanglement entropy can be computed in the same formulation. 

For simplicity, we assume that the classical model is defined on the square
lattice, 
and a spin-$S$ degrees of freedom lives on every site. 
We again assume that the energy $E(c)$ consists only of interactions between
nearest-neighbor spins.
As illustrated in Fig.~\ref{fig:AB_aiib}, each spin is {\it duplicated} and
an infinitely strong ``ferromagnetic'' interaction is added 
so that the two copies of the original spin are always in the same state 
(no spurious degrees of  freedom are introduced). 
Then the two regions $A$ and $B$ are introduced in such a way that 
all the spin-spin interactions in $E(c)$ take place inside $A$ or $B$.
In other words, the only allowed couplings between $A$ and $B$ are the
infinitely strong ``ferromagnetic'' interactions 
between copies of the same physical spin. 
If one prefers to think in terms of the original spins only (not duplicated), 
this amounts to saying that the regions $A$ and $B$ are  overlapping around
their boundary. 

In this setup, each state $|c\rangle$ can be labeled in the following way:
\begin{equation}
 |c\rangle= |a,i\rangle \otimes |b,i\rangle.
	\label{eq:abc}
\end{equation}
Here, each original spin lying at the boundary is effectively ``split'' and has
one copy in $|a,i\rangle$ and the other in $|b,i\rangle$.
The Schmidt decomposition \eqref{eq:schmidt} is then constructed 
using the following states:
\begin{subequations}
\begin{eqnarray}
 |{\rm RK}_i^\Omega \ra &:=& \frac1{\sqrt{\mathcal{Z}_i^\Omega}} 
 \sum_\omega e^{-\frac{1}{2} E_\Omega(\omega,i)}|\omega,i\ra, 
   \label{eq:RK_Omega} \\
 \text{with} &~&  
 \mathcal{Z}_i^{\Omega}=\sum_\omega e^{-E_\Omega(\omega,i)} 
 ~~(\Omega=A,B).
\end{eqnarray}
\end{subequations}
Here, the difference from Eq.~\eqref{eq:RK_region} 
is the presence of $i$ inside the ket $|\omega,i\ra$ for both $\Omega=A$ and
$B$, 
which ensures the mutual orthogonality of $\{ |{\rm RK}_i^\Omega \ra\}$.

\subsection{Transfer matrix calculation of the reduced density matrix spectrum}
\label{sec:transfer_matrix}

In the previous subsection, 
the spectrum $\{p_i\}$ of the reduced density matrix 
has been expressed in terms of the classical partition functions 
with spins fixed in the boundary region.
In the cylindrical geometry of Fig.~\ref{fig:AB} 
with circumference $L$ and height $2h$, 
we can relate this spectrum to the ground state of a 1D quantum spin model 
using the transfer matrix formalism.

Corresponding to the 2D classical model, 
we introduce the transfer matrix $\mathcal{T}$ in the upward direction 
in such a way that it connects spin configurations 
on neighboring ``rings'' winding around the cylinder
(e.g., $a_1$ and $a_2$ shown in Figs.~\ref{fig:AB_aib} and \ref{fig:AB_aiib}).
The classical partition function \eqref{eq:Zcl} is then expressed as 
\begin{eqnarray}
 \mathcal{Z} &=& 
 \sum_{ a_{h-1},\dots,a_1}
 \sum_i
 \sum_{ b_1,\dots,b_{h-1}}\\
 &&\la a_h| \mathcal{T} |a_{h-1}\ra \dots
 \la a_2| \mathcal{T} |a_1\ra  
 \la a_1| \mathcal{T} |i\ra \\
 &&\times \la i| \mathcal{T} |b_1\ra 
 \la b_1| \mathcal{T} |b_2\ra \dots
 \la b_{h-1}| \mathcal{T} |b_h\ra \\
 &=& \la a_h | \mathcal{T}^{2h} | b_h\ra.
\end{eqnarray}
Here, the spin configurations, $a_h$ and $b_h$, at the top and bottom edges of
the cylinder 
are fixed. 
In this setup, the classical probability to find a given configuration
$i$ on the ring (boundary) is 
\begin{eqnarray}
 p_i=\frac{1}{\mathcal Z} \langle a_h | \mathcal{T}^{h} |i\rangle \langle i |
\mathcal{T}^{h} |b_h \rangle.
\end{eqnarray}

We now consider the limit of a long cylinder $h\gg L$ so that 
only the dominant eigenvector $|g\ra$ of $\mathcal{T}$ 
(with the largest eigenvalue $m_0$) contributes. 
Using $\mathcal{T}^{h}\simeq m_0^{h}|g\rangle\langle g|$, we get:
\begin{eqnarray}
 p_i \simeq  | \langle i | g\rangle |^2.
	\label{eq:pi}
\end{eqnarray}
If the transfer matrix is related to 1D quantum Hamiltonian $\mathcal{H}$ 
via $\mathcal{T} \simeq e^{- \tau \mathcal{H}}$ (with $\tau$ being a small time
interval), 
$|g\ra$ is the ground state of $\mathcal{H}$.
Then, Eq.~\eqref{eq:pi} means that the complete spectrum of the reduced density
matrix $\rho_A$ is given 
by the elements of the ground state vector $|g\ra$. 
In spite of its simplicity, to our knowledge, this result has not been reported
previously in the literature.

In the rest of the paper, 
we will make an extensive use of this property to calculate the entanglement
entropy \eqref{eq:S_RK}. 
We study several RK states (dimer, vertex, and Ising models) 
defined on the infinite cylinder 
for relatively large values of $L$
(maximally, $L=48$ for the dimer models and $L=32$ in the six-vertex models with no field). 

\subsection{Thermodynamic extension}
\label{sec:thermo_extension}
The spectrum $\{ p_i \}$ of the reduced density matrix $\rho_A$ for a RK state 
has a simple interpretation in terms of boundary free energy of the classical
model.
Using Eq.~\eqref{eq:schmidt}, we get
\begin{equation}
	-\log p_i= F_i^A+F_i^B-F,
\label{tmfreenrj}
\end{equation}
where $F_i^A := -\log \mathcal{Z}_i^A$ (resp.\ $F_i^B$) 
is the free energy of the subsystem $A$ (resp.\ $B$) 
with its boundary with $B$ (resp.\ $A$) fixed in the configuration $i$.  
Also, $F=-\log \mathcal{Z}$ is the free energy of the whole system, 
without any constraint on the spin configuration at the boundary between $A$ and $B$. 
We now identify the r.h.s.\ of Eq.~\eqref{tmfreenrj} as an effective energy $2E(i)$ for the boundary spins. 
\footnote{
Here the factor $2$ is inserted because it will be convenient in Sec.~\ref{sec:RKtoGaudin} 
when identifying $E(i)$ as a Coulomb gas model. 
One may also think that the interactions occuring in $A$ and $B$ make a contribution $E(i)$ each 
and that the total contribution is twice of it.
}
Then the entanglement entropy $S^{\rm VN}(A)$ in Eq.~\eqref{eq:S_RK} 
can be interpreted as the ``thermal'' entropy for the boundary spins. 
We push further this thermodynamic interpretation of $S^{\rm VN}(A)$ 
by introducing a parameter $\beta$ in order to modify the probabilities $p_i$:
\begin{equation}
 p_i=e^{-2E(i)} \to p_i(\beta)=\frac{1}{Z(\beta)} e^{-\beta E(i)},
\label{eq:p_i_beta}
\end{equation}
where $Z(\beta)=\sum_i e^{-\beta E(i)}$ is a normalization factor 
(with $Z(\beta=2)=1$). 
Here, $\beta$ plays the role of an effective inverse temperature for the boundary spins 
(but a priori {\it not} for the bulk of the classical system defined by Eq.~\eqref{eq:Zcl}). 
This allows us to generalize the entropy $S(\beta=2)=S^{\rm VN}(A)$ to arbitrary $\beta>0$. 
It can be computed through the standard thermodynamical relation 
$S(\beta)=(1-\beta \partial_\beta)Z(\beta)$. 
This formulation will be useful in Sec.~\ref{sec:RKtoGaudin}. 
We note that a similar extension of entanglement entropy has also been discussed by Li and Haldane.
\cite{Haldane_talk,lh08}

\section{{From critical dimer RK states to Dyson-Gaudin coulomb gas}}
\label{sec:RKtoGaudin}

In this section, we study critical dimer RK states on bipartite lattices 
and a related 1D classical gas model. 
Using the formulation of Sec.~\ref{sec:RK_states}, 
we compute the entanglement entropies of the dimer RK states 
using the dominant eigenvectors of the transfer matrices. 
As described in Appendices~\ref{sec:hexa_tm} and \ref{sec:square_tm}, 
the transfer matrices of the dimer models can be expressed as free fermion Hamiltonians, 
and their dominant eigenvectors are Slater determinants. 
In particular, for the hexagonal lattice dimer model, 
the resulting probabilities $\{p_i\}$ coincides with 
the Boltzmann weights of a 1D lattice gas 
interacting via a repulsive logarithmic potential (Dyson-Gaudin gas) 
at the inverse temperature $\beta=2$. 
This gas model is also related to the discretized\footnote{The particle coordinates are restricted to be multiple of $2\pi/L$} ground state wave function 
of the Calogero-Sutherland model. 
Therefore, the entanglement entropy of the dimer RK state 
and the Shannon entropy of the discretized Calogero-Sutherland ground-state wave function   
coincide with the thermal entropy of the gas model. 
As we will demonstrate in Secs.~\ref{sec:dimers_hexa} and \ref{sec:gaudin}, 
this entropy contains a non-extensive constant contribution, Eq.~\eqref{eq:S0}. 
In Sec.~\ref{sec:field}, we derive it analytically 
using the free field representation of the gas model in the continuum limit.

\subsection{Critical dimer RK states}
\label{sec:dimers_hexa}

We start from the RK states \eqref{eq:RK} 
constructed from the dimer models on the hexagonal and square lattices. 
The energy $E(c)$ takes zero for any fully packed dimer configuration $c$ 
and infinity otherwise. 
These RK states are the ground states of quantum dimer models 
at special points 
and are know to be critical.\cite{rk88,msc01} 
By associating an Ising variable $\sigma_j=\pm 1$ with the dimer occupation on each bond $j$, 
we can adopt the formulation presented 
in Secs.~\ref{sec:schmidt1} and \ref{sec:transfer_matrix} 
to compute the half-cylinder entanglement entropy $S^{\rm VN}$.

As described in Appendices~\ref{sec:hexa_tm} and \ref{sec:square_tm}, 
the transfer matrices ${\cal T}$ of the dimer models are expressed as free fermion Hamiltonians.   
Their dominant eigenvectors $|g\ra$ are Slater determinants (Fermi sea), 
and the weight $p_i= |\la i|g \ra|^2$ can be computed by evaluating a determinant. 
Let us focus on the hexagonal case, where 
a simple expression for $\{p_i\}$ is available. 
We here assume that $L$ is a multiple of $6$ for the sake of simplicity. 
A generic configuration $i$ of the boundary will be given by a number $n \in \{0,\ldots,L\}$ of fermions 
living on the vertical edges of the boundary and their positions:
\begin{equation}
 0\leq \alpha_1<\ldots<\alpha_n\leq L-1.
\end{equation}
It is also shown in Appendix \ref{sec:hexa_tm} that, 
in the limit $h\gg L$, 
the only nonzero probabilities correspond to $n=2L/3$ fermions
and are given by a Vandermonde determinant, which simplifies into:
\begin{equation}\label{eq:proba_hexa}
 p_i=\frac{1}{L^n}\prod_{1\leq j<j'\leq n} 4\sin^2 \left(\frac{\pi}{L}(\alpha_j-\alpha_{j'})\right).
\end{equation}
This equation is invariant under $n\rightarrow L-n$. Therefore it is easier to compute it with $n=L/3$ fermions, instead of $n=2L/3$.  
As we will see in the next section, 
these probabilities coincide with the Boltzmann weights of a 1D lattice gas. 
Note that the present calculation is done 
without any constraint on the configurations, $a_h$ and $b_h$, at the top and bottom of the cylinder. 
The number $n$ of fermions leading to nonzero probabilities can be controlled 
by imposing certain boundary conditions. 

\begin{figure}
\begin{center}
\includegraphics[angle=-90,width=8.5cm]{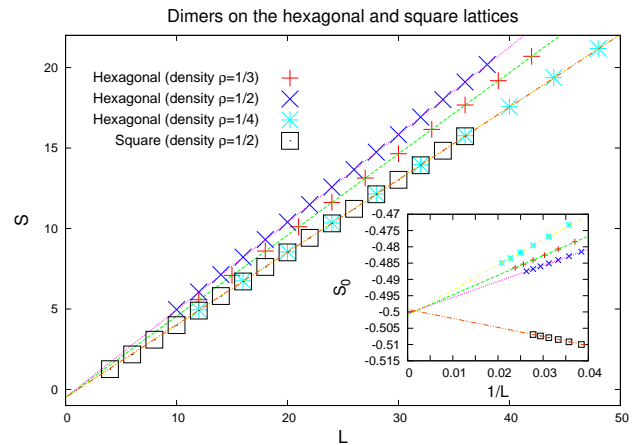}
\caption[1]{(color online)
Entanglement entropy of RK states corresponding to 
dimer models on the hexagonal lattice (with fermion density $\rho=1/4,~1/3,~1/2$) 
and the square lattice (with $\rho=1/2$). 
In all cases, the entropy scales as $S\simeq \alpha L+S_0+b/L$ with $S_0=-0.500(2)$.
The inset shows $S'_0$, the subleading constant computed this time from a linear fit $S\simeq a'L+S'_0$ on the 
interval $[12,L]$ as a function of $1/L$. The convergence towards $-0.500(1)$ can be seen.
}\label{fig:hex_square}
\end{center}
\end{figure}

Figure~\ref{fig:hex_square} shows the scaling of the entanglement entropy $S(L)$ 
for both the hexagonal and square cases. 
Here, in the square case, the weight $p_i$ is computed 
by numerically evaluating the determinant in Eq.~\eqref{eq:square_det}. 
In the hexagonal case, we examine different fermion densities $\rho=n/L$ 
(or equivalently flux/winding sectors in the dimer language). 
In every case, the scaling of $S(L)$ appears to be approximately linear in $L$. 
The slope is non-universal, as expected, and depends on the details of the system. 
The most interesting result is the existence of a finite constant $S_0$. 
By fitting the large-$L$ values by $S(L)=\alpha L+S_0+b/L$, 
we find $S_0 = -0.500\pm 0.002$ in all cases. 
These results suggest that $S_0=-1/2$ is a universal number for this family of RK wave functions.

\subsection{Dyson-Gaudin gas on a circle}
\label{sec:gaudin}

\begin{figure}
\begin{center}
\includegraphics[width=7cm]{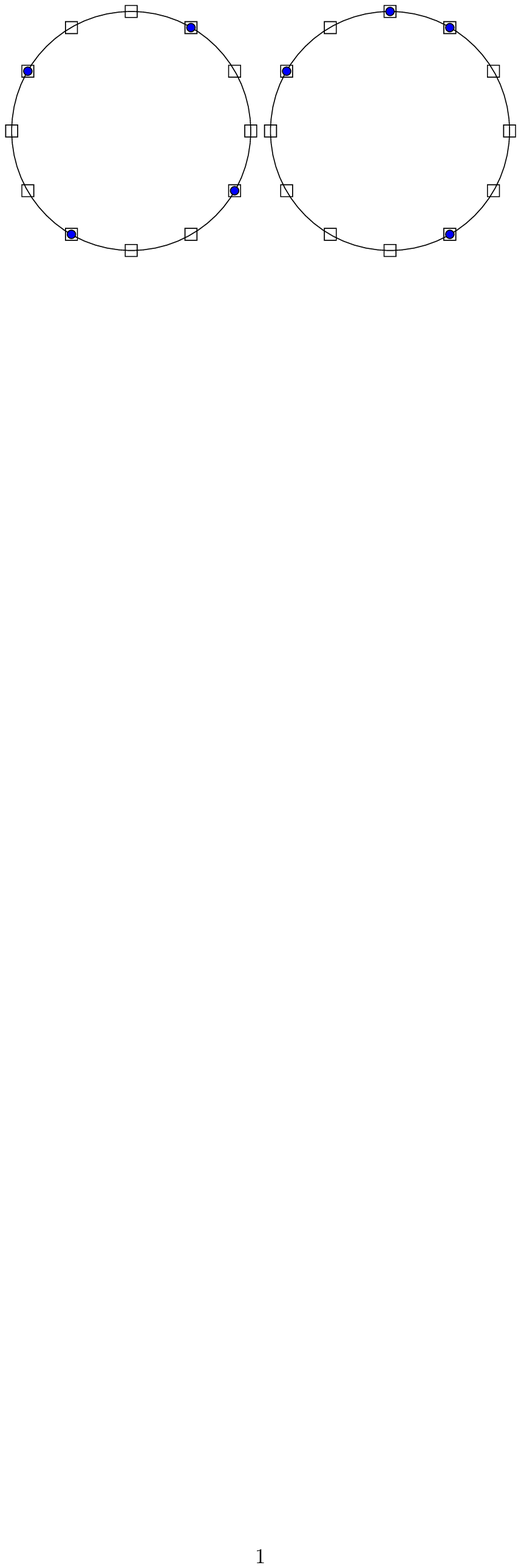}
\caption[1]{
System of $n=4$ charges $Q=-1$ on a circle with $L=12$ sites. 
The left shows the configuration with highest probability (degeneracy: $3$)
and the right shows a random configuration. 
On each site, a ``background'' charge $Q=+1/2$ is  added.
}\label{fig:Coulomb_gas}
\end{center}
\end{figure}

We consider a system of $n$ charges $Q=-1$, 
living on a periodic one-dimensional lattice with $L$ sites as in Fig.~\ref{fig:Coulomb_gas}. 
These charges interact via a 2D Coulomb repulsive potential equal to minus the logarithm of their mutual distance. 
For convenience, we add a uniform background of $L$ charges $Q=+1/2$ located on each site of the lattice. 
A configuration of the system will be determined 
by the positions $1\le \alpha_1 < \alpha_2 < \ldots <\alpha_{n}\le L$ of the charges. 
Two charges cannot occupy the same site. 
The energy of a given configuration  $i$ is:
\begin{eqnarray}\label{eq:energy_gaudin}
 E(i) = -  {\!\!\!\!} \sum_{1\leq j<j'\leq n} {\!\!\!\!}
         \log \left|e^{\frac{2\pi i}{L}\alpha_j}-e^{\frac{2\pi i}{L}\alpha_{j'}}\right|+\frac{n}{2}\log L,
\end{eqnarray}
where the second term comes from the interaction of each charge with the background. 
This may be viewed as a discretized version of the Dyson gas,\cite{dyson} 
and has been studied by Gaudin\cite{gaudin} (hence we name it the Dyson-Gaudin gas). 
As in Sec.~\ref{sec:dimers_hexa}, the ``particle-hole'' symmetry $n\rightarrow L-n$ is worth noticing. 
The classical 1D partition function is given by $Z(\beta)=\sum_i e^{-\beta E(i)}$ 
for an inverse temperature $\beta$, 
and the classical entropy can be calculated from it: 
\begin{equation}\label{eq:S-Z}
 S(\beta)=\left(1-\beta \frac{\partial}{\partial \beta}\right)\log Z(\beta).
\end{equation}
Gaudin\cite{gaudin} has evaluated the partition function exactly 
in the special cases where $\beta=2\lambda$ with $\lambda$ an integer and $\lambda<L/(n-1)$: 
\begin{equation}
 Z_n^{(L)}(2\lambda)=\frac{(n\lambda)!}{n!L^{n(\lambda-1)}(\lambda !)^n}.
 \label {01-gaudin}
\end{equation}

The probabilities $\{p_i\}$ in Eq.~\eqref{eq:proba_hexa} 
calculated for the hexagonal lattice dimer model 
coincides exactly with the Boltzmann weights of the Dyson-Gaudin gas model with $\beta=2$: $p_i= e^{-2E (i)}.$

Hence the entanglement entropy for this dimer model 
corresponds to the thermal entropy $S(\beta=2)$ of the gas model. 
Note that $Z(\beta=2)=1$ if we set $\lambda=1$ in Eq.~\eqref{01-gaudin}  
because $p_i$ in Eq.~\eqref{eq:proba_hexa} are already normalized. 
In the spirit of Sec.~\ref{sec:thermo_extension}, 
we can generalize the entanglement entropy 
by changing the inverse temperature $\beta$. We define 
$p_i(\beta)
=p_i^{\beta/2}/Z(\beta)
=e^{-\beta E_i}/Z(\beta)$, and the associated Shannon entropy 
\begin{equation}
 S(\beta)=-\sum_i p_i(\beta)\log p_i(\beta), 
\end{equation}
coincides with the thermal entropy \eqref{eq:S-Z} of the gas model. 
Notice that Gaudin's solution~\eqref{01-gaudin} cannot be used 
to compute the entropy $S(\beta)$ because it is valid only for special values of $\beta$. 
Instead, we compute $S(\beta)$ numerically 
by explicitly summing over all the configurations.

Here we mention the connections with other models. 
The Dyson gas model emerges in the weights $|\la i|g \ra|^2$ of 
the Jastrow-type ground-state wave function of the Calogero-Sutherland (CS) model.\cite{Suth1,Suth2} 
The inverse temperature $\beta$ of the gas model 
is related to the coupling constant of the CS model. 
The CS model is described as a Tomonaga-Luttinger liquid at low energies. 
According to a spectral analysis of the CS model,\cite{bosonization_ll} 
the boson radius $R$ is related to $\beta$ as
\begin{equation}\label{eq:R-beta}
  R=\sqrt{\frac\beta2}. 
\end{equation}
This allows us to control $R$ simply by changing $\beta$. 
This relation will also be justified from a different viewpoint 
in the next section. 
The Haldane-Shastry model\cite{Haldane88,Shastry88} is a discretized version of the CS model,  
and its Jastrow ground state (Gutzwiller-projected Fermi wave function) 
has the weights exactly obeying the Dyson-Gaudin gas with $\beta=4$. 
This model has $R=\sqrt{2}$ because of the $SU(2)$ symmetry, 
in consistency with Eq.~\eqref{eq:R-beta}.
Note also that the same wave function is known to be an extremely good ansatz 
for the ground state of the Heisenberg chain.

\begin{figure}
 \begin{center}
  \includegraphics[angle=-90,width=8cm]{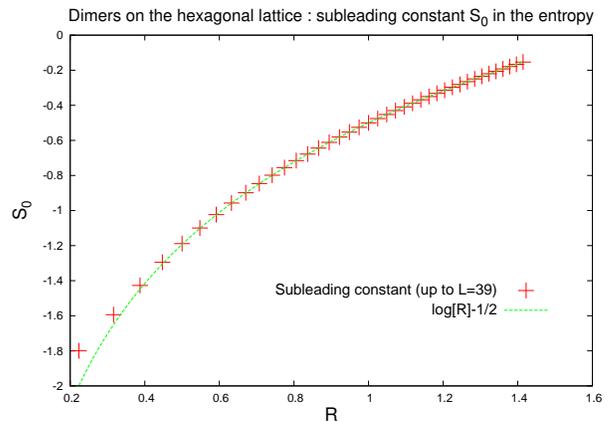}
\caption[1]{(color online)
The subleading constant $S_0$ in the thermal entropy of the Dyson-Gaudin gas 
with density $\rho=1/3$, as a function of $R=\sqrt{\beta/2}$. 
At $\beta=2$, this coincides with the constant part 
in the entanglement entropy of the hexagonal dimer RK state shown in Fig.~\ref{fig:hex_square}. 
The data well obeys Eq.~\eqref{eq:S0_R}. 
}\label{fig:hex_beta}
 \end{center}
\end{figure}

Now we analyze the thermal entropy $S(\beta)$ of the gas model. 
We extract the non-extensive constant contribution $S_0$ 
as in Fig.~\ref{fig:hex_square}, 
and plot it as a function of $R=\sqrt{\beta/2}$ in Fig.~\ref{fig:hex_beta}. 
We find that the data agrees well with a simple relation 
\begin{equation}\label{eq:S0_R}
 S_0=\log R-\frac12.
\end{equation}
This expression is derived analytically in the next section. 
It should be noted that the subleading constant \eqref{eq:S0_R} is increasing with $\beta$, contrary to the total entropy, 
which is decreasing with $\beta$, as should be the case in classical thermodynamics.

\subsection{Free bosonic field}
\label{sec:field}

In this section, we obtain the expression of the entropy $S_0$ 
using a field theoretical approach. 
Our goal is to obtain a continuous expression 
for the partition function of the gas studied in the last section 
and deduce from it the expression of the entropy.

We consider a continuous distribution, $\rho (\theta),$ of electric charges on the unit circle.
The expression of the electrostatic energy Eq.~\eqref{eq:energy_gaudin} is given by:
\begin{eqnarray}
 E= -\frac{1}{2}
    \int_{0}^{2\pi} \!\!\! \rho(\theta )d\theta
    \int_{0}^{2\pi} \!\!\! \rho(\theta')d\theta' 
    \log \bigg |2\sin \left( \frac{\theta-\theta'}{2} \right) \bigg| .
 \label {bolz}
\end{eqnarray}
We define a field $\phi(\theta)$ measuring the amount of charge in the interval $[0,\theta]$ in units of $2\pi $:
\begin{eqnarray}\label {phi-def}
 \phi(\theta)=2\pi \int_0^\theta \rho(\sigma)d\sigma 
\end{eqnarray}

By performing partial integrations twice, 
the energy is rewritten as
\begin{eqnarray}
E[\phi]=\frac{1}{64 \pi^2 } \int_{0}^{2\pi}d\theta \int_{0}^{2\pi} d\theta' \  \left (\frac{\phi(\theta)-\phi(\theta')}{\sin(\frac{\theta-\theta'}{2})} \right )^2.
\label {bolz1}
\end{eqnarray}

Since the functional integration is over $\rho=\phi^\prime$, 
the zero mode of $\phi$ is unphysical and should be discarded 
(it can be removed by adding an appropriate constant to Eq.~\eqref{phi-def}).  
By expanding the field $\phi$ over modes
\begin{eqnarray}
\phi=2\pi\sum_{m\ge 1} \left( x_m e^{im\theta}+\bar x_m e^{-im\theta} \right), 
 \label {modes-phi}
\end{eqnarray}
the energy Eq.~\eqref{bolz1} reduces to 
\begin{eqnarray}
 E=\frac{(2\pi)^2}{2}\sum_{m\ge 1} {m|x_m|^2}.
 \label {02}
\end{eqnarray}
Integrating the Boltzmann weight $e^{-\beta E}$ over the modes $dx_md\bar x_m,$ 
we obtain the partition function $Z_{\rm sphere}$ of the gas:
\begin{eqnarray}
Z_{\rm sphere}=\prod_{m\ge 1} \frac{2}{2\pi \beta m  }.
 \label {Z-zeta}
\end{eqnarray}
To find possible universal contributions to the corresponding free energy, 
this expression of course needs to be regularized. 
Following Nahm,\cite{Nahm} we regularize the measure  $dx_md\bar x_m$ 
to take into account the finiteness of the number of states. 
We set $|x_m|=\rho_m$ and take the measure to be $d(2\pi [\rho_m^2]^{f(m/\Lambda)})$ 
where $f(x)=1$ in the interval $[0,1]$ and decreases to $f(\infty)=0$ sufficiently fast. We obtain :
\begin{equation}
 Z_{\textrm{sphere}}=\prod_{m=1}^\infty \left(\frac{1}{\pi m \beta}\right)^{f(m/\Lambda)} \Gamma(1+f(m/\Lambda))
\end{equation}
The Euler-MacLaurin formula yields :
\begin{equation}\nonumber
\prod_{m=1}^\infty\!\!\left(\!\frac{1}{\pi \beta}\!\right)^{\!\!f(\frac{m}{\Lambda})}\!\!\!\Gamma\!\left(1\!+\!f(\frac{m}{\Lambda})\right)=
\!\sqrt{\pi \beta}e^{\textstyle{\Lambda \!\int_{0}^\infty \!\!g_\beta(x)dx}}(1+o(1))
\end{equation}
\begin{equation}
g_\beta(x)=\log \Gamma(1+f(x))-\log (\pi \beta)f(x)
\end{equation}
Here we decompose $f$ into the sum of a step function on $[0,1]$ and a function vanishing on $[0,1]$ :
\begin{equation}
 \prod_{m=1}^\infty \left(\frac{1}{m}\right)^{f(\frac{m}{\Lambda})}=\frac{1}{\sqrt{2\pi}}e^{\textstyle{-\Lambda\int_0^\infty \log (\Lambda x)f(x)dx}}(1+o(1)),
\end{equation}
where we have used the Stirling formula on the term corresponding to the step function, and Euler-MacLaurin's on the other one. Putting everything together :
\begin{equation}
 Z_{\textrm{sphere}}=\sqrt{\frac{\beta}{2}}e^{\Lambda \int_0^\infty dx\log \Gamma(1+f(x))-\log(\pi \beta \Lambda x)f(x)}(1+o(1))
\end{equation}
The regularized partition function is obtained after removing the exponential factor, $\Lambda^{-a\Lambda}\mu^\Lambda$, which can be thought as the ``extensive'' part
\begin{eqnarray}
Z_{\rm sphere}=\sqrt{\frac{\beta}{ 2}}.
 \label {Zc0}
\end{eqnarray}
This result is simply equal to the $\zeta$ regularization of Eq.~\eqref{Z-zeta}.
In this derivation, the normalization of the integration measure over the modes is adjusted so that $Z_{\rm sphere}=1$ at $\beta=2$, to agree with our microscopic definition of the partition function, $Z_{\rm sphere}(\beta)=\sum_{i}p_i^{\beta/2}$, in the discrete model. 
From this expression, we deduce the thermal entropy:
\begin{eqnarray}
S_0=\frac{1}{2}\log \left(\frac{\beta}{2}\right)-\frac{1}{2} 
 \label {entropie1}
\end{eqnarray}
in agreement with our numerical result in Fig.~\ref{fig:hex_beta}. 

Alternatively, we can  normalize the field $\phi$ differently 
so as to include the inverse temperature $\beta$ in its definition. 
We introduce a radius $R$ and set $\beta=2 R^2$. 
The field $\phi$ is now defined modulo $2\pi R$.
We can extend its range of definition 
by requiring it to be a harmonic function on the unit disk $\Omega,$ 
and express the energy as a Dirichlet integral:
\begin{eqnarray}
\frac{\beta}{2}{E}=\frac{1}{4 \pi} \int\int_{\Omega} dzd\bar z\  \partial_z\phi\partial_{\bar z}\phi,
 \label {Dirichlet}
\end{eqnarray}
where $z=x+iy$. 
This coincides with the action~\eqref{eq:FreeFieldAction} 
except that the range of integration is limited to the unit disk $\Omega$. 

One can view Eq.~\eqref{Dirichlet} as the action of a closed string\cite{polchinski} 
propagating on a circle of radius $R$ with a Regge slope $\alpha'=2$. 
Now, if we include the center of mass (zero mode) into the definition of $\phi$, 
we see that the regularized measure $[d\phi]$ becomes \textit{invariant} under rescaling of the field, 
so that we can also take the field defining the measure to be defined modulo $2\pi R$. 
The partition function $Z_{\rm sphere}$ of the electrostatic gas is obtained 
by sewing together two disks to form a sphere 
and is given by the partition function of a closed string propagating on a circle of radius $R$.
Proceeding in this way, the oscillators do not contribute and it reduces to the center of mass integral:
\begin{eqnarray}
 Z_{\rm sphere}=R,
  \label {Zc}
\end{eqnarray}
in agreement with Eq.~\eqref{Zc0}.

To understand this result, consider a closed string $\phi(\sigma,T)$ propagating in the Euclidean time $T$. 
Its partition function on the cylinder $[0,2\pi ]\times [0,T]$ with boundary fields equal to $\phi_{1,2}$ 
defines the propagator $Z(\phi_1,\phi_2)$. 
We evaluate the torus partition function $Z_{\rm torus}$ 
by taking the trace of the propagator over $\phi=\phi_1=\phi_2$.
\begin{eqnarray}
 Z_{\rm torus}=\int [d\phi] Z(\phi,\phi).
 \label {Z-torus}
\end{eqnarray}
If we decompose the field $\phi$ into the sum of a harmonic function 
equal to $\phi_1,\ \phi_2$ at the two boundaries and a field vanishing at the boundaries, 
we  factorize the propagator into two pieces: 
a classical one equal to $e^{-\frac{\beta}{2}E(\phi_1)-\frac{\beta}{2}E(\phi_2)}$ in the limit of large $T$, 
and the partition function $Z_{00}$ with Dirichlet boundary conditions. 
Thus, In the limit of large $T$, $Z_{\rm torus}=Z_{\rm sphere}Z_{00}$. 
But $Z_{00}/Z_{\rm torus}=1/R$ is the stationary probability distribution 
of the center of mass of the string diffusing on a circle of radius $R$ 
and the result Eq.~\eqref{Zc} follows.
This formulation has the advantage to be easily generalizable 
to a closed surface of characteristic $\chi$: 
\begin{equation}\label{eq:Euler_characteristic}
 Z_{\chi}=R^{\chi/2}.
\end{equation}
We expect this formula also to apply to the case of open boundary conditions with suitable chiral boundary conditions. 

Let us mention that Gaudin's partition function~\eqref{01-gaudin} 
has the same asymptotic expression as $Z_{\rm sphere}$ in Eq.~\eqref{Z-zeta} 
if we remove the non-universal extensive part. 
Indeed, by taking $n\to\infty$ while keeping $\rho=n/L$ constant, 
the Stirling's formula $n! \simeq \sqrt{2\pi n}~(n/e)^n$ applied to Eq.~\eqref{01-gaudin} gives:
\begin{eqnarray}
Z_{n}^{(L)}(2\lambda)=\mu(\lambda)^n R,~~
\text{with}~~ \mu(\lambda)=\frac{\rho^{{\lambda}-1}{\lambda }^{\lambda}{e }^{\lambda-1}}{\lambda !}.
 \label {gaudin-stirling}
\end{eqnarray}
Gaudin's formula~\eqref{01-gaudin} is only valid at integer values of $\lambda$ and we cannot use it to evaluate the entropy. However, the
denominator factor $L^{(\lambda-1)n}$ in Eq.~\eqref{01-gaudin} may be viewed as a regularization of the Dyson partition function,\cite{dyson,mehta}  
where $L$ plays the same role as $\Lambda$ defined above, and the entropy also derives from the Dyson gas partition function.

The compactification radius of the (continuum) CS ground state\cite{Suth1,Suth2}
is known to be $R=\sqrt{(\lambda=\beta/2)}$ from Ref.~\onlinecite{bosonization_ll}.
On the other hand, our numerics on the {\it discretized} version of the CS wave function indicate
that the entropy constant is equal (within our numerical accuracy) to
$S_0=\log(\sqrt{\beta/2})-1/2$. From the analytical derivations presented above, this entropy constant must be related
to the boson radius. We therefore conclude that the {\it discretized} CS state has long distance properties described
by the same boson compactification radius as the original continuum CS wave function. It is interesting to notice that this identification 
has been made through the ground-state structure of the CS model, {\it i.e.} without relying on the spectral properties unlike 
preceding approaches.\cite{bosonization_ll} 

\subsection{Phase transition toward a crystal state}
\label{sec:transition}

\begin{figure}
 \begin{center}
  \includegraphics[angle=-90,width=8.5cm]{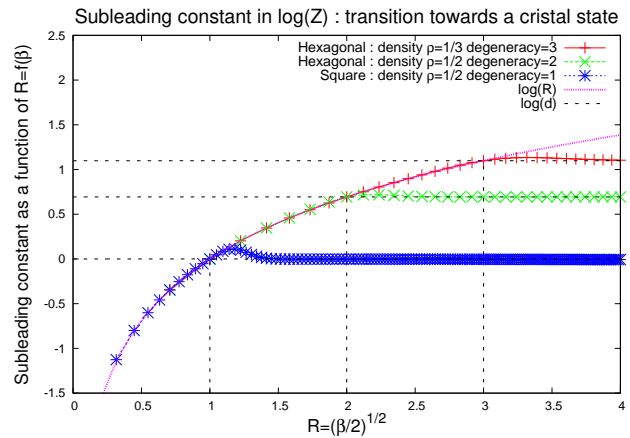}
\caption{(color online) 
Liquid-crystal transitions induced by the boundary temperature $\beta^{-1}$ 
in the dimer models. 
Here, the subleading constant $C$ in $\log Z$ is shown. 
The critical radius is $R_c=d$ ($\beta_c=2d^2$), 
where $d$ is the degeneracy of the ground state. 
The constant $C$ is expected to obey Eq.~\eqref{eq:subleading_c}. 
The discrepancy slightly after $R_c$ is very likely due to finite-size effects.
}\label{fig:phase_transition}
 \end{center}
\end{figure}

The numerical results of Fig.~\ref{fig:hex_beta} show 
that for not too large values of $\beta$, 
the non-extensive contribution to the entropy is given by $S_0=\log R-1/2$. 
As should be clear from the previous section, 
this can only be true if the system is described by a massless (but compactified) free field. 
But for sufficiently large $\beta$, 
the system undergoes a transition to a crystal state 
with spontaneous translation symmetry breaking. 
A simple way to understand that such a crystal is expected at large $\beta$ is 
to notice that for $\beta\to \infty$, 
only the particle configurations $i$ for which the (original) probability $p_i$ is maximum survive. 
For a particle density $1/d$, 
this selects $d$ periodic configurations with equally spaced particles 
(see the left panel in Fig.~\ref{fig:Coulomb_gas}). 
Adding fluctuations around these regular configurations will add {\it extensive} contributions to $S(L)$, 
while keeping the subleading constant
\begin{equation}
 S_0=\log d
\end{equation}
stable in a crystal phase with a d-fold spontaneous symmetry breaking. 
Due to some finite-size effects, 
it turns out that the liquid-crystal transition is easier to see 
in the non-extensive part $C$ of $\log Z$ (rather than that of $S$). 
The data displayed in Fig.~\ref{fig:phase_transition} are consistent with
\begin{equation}\label{eq:subleading_c}
 C=\left\{\begin{array}{ccc}\log R,& &R\leq R_c=d \\ \log d,& & R\geq R_c=d.\end{array}\right.
\end{equation}
Using $S_0=\left(1-\frac{R}{2}\partial_R\right) C$, 
we can recover the subleading term $S_0$ in the entropy:
\begin{equation}
 S_0=\left\{\begin{array}{ccc}\log R-1/2&,&R<R_c=d \\ \log d&,& R> R_c=d.\end{array}\right.
\end{equation}
It should be noted that the transition is only visible on the subleading terms of $S$ and $\log Z$.

The crystallization can also be understood from a free field point of view. 
Let us perturb the action~\eqref{eq:FreeFieldAction} by a d-fold symmetry breaking \textit{boundary} field :
\begin{eqnarray}
Z_{h_d}=\left\langle e^{h_d\int_{0}^{2\pi} \cos ({d\phi}) d\theta}\right\rangle_{\textrm{sphere}},
 \label {perturb}
\end{eqnarray}
where the integral is taken over the equator of the sphere. In a spin wave approximation, the anomalous dimension of the field $h_d$ is:
\begin{eqnarray}
x_{d}=1-\frac{d^2}{R^2}.
 \label {dimention_sw}
\end{eqnarray}
Thus, the perturbation becomes relevant when $R\ge d$ 
in agreement with our observations (Fig.~\ref{fig:phase_transition}).

This transition is completely analogous to the localization transition 
of a macroscopic degree of freedom coupled to a dissipative environment in presence of a periodic potential.\cite{Schmid} 
In this context, the inverse temperature $\beta=2R^2$ is a friction coefficient. 
A similar transition is observed in the XXZ chain (see the following section), 
but at a {\it different} value of the compactification radius ($R=\sqrt{2}$), 
compatible with the bulk roughening transition of the 6-vertex model.

\section{Spin-$1/2$ XXZ chain and six-vertex RK states}
\label{sec:xxz_chain}

In this section, we consider the Shannon entropy \eqref{eq:Sdef} 
defined from the ground state $|\psi\rangle$ of the spin-$\frac{1}{2}$ XXZ
chain:
\begin{equation}\label{eq:XXZ}
 \mathcal{H}=\sum_j \left(
	\sigma^x_j \sigma^x_{j+1} + \sigma^y_j \sigma^y_{j+1}+\Delta  \sigma^z_j
\sigma^z_{j+1}\right)
    -h \sum_j \sigma^z_j.
\end{equation}
This Hamiltonian is related to the transfer matrix of the classical six-vertex
model 
on the square lattice.\cite{Baxter82,Kasteleyn75}
Thus, using the argument of Sec.~\ref{sec:RK_states}, the Shannon entropy $S$
here 
can also be interpreted as the entanglement entropy of the RK state built from
this vertex model. 
Since the magnetization per site $M=\frac{1}{L}\sum_i \sigma_i^z$ is a conserved
quantity, 
we can work in a sector with fixed $M$. 
We calculate the ground state of $\mathcal{H}$ for finite periodic chains 
using Lanczos diagonalization (up to $L=32$ for $M=0$ and $L=40$ for $M=1/2$), 
and evaluate the Shannon entropy $S$ from it.

\begin{figure}
\begin{center}
\includegraphics[width=6cm,angle=-90]{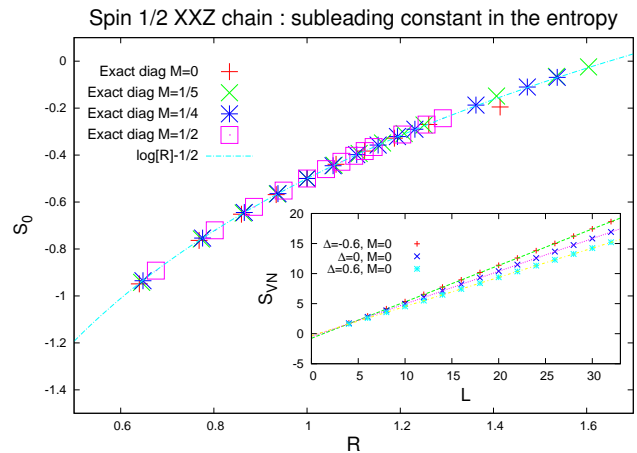}
\caption[1]{(color online) 
The subleading constant $S_0$ in the entropy $S$ 
extracted from the critical ground state of the XXZ chain \eqref{eq:XXZ}.  
The examined values of $\Delta$ 
range from $-0.8$ to $1$ for $M=0$ 
and from $-0.8$ to $8$ for $M=1/5,~1/4,$ and $1/2$. 
The inset shows the fitting of the data with the scaling form $S = \alpha L+S_0 +
b/L$. 
The constant $S_0$ well obeys the proposed universal formula $\log R
-\frac{1}{2}$. 
Close to the isotropic point ($\Delta=1$ and $M=0$) with $R=\sqrt{2}$, 
a small discrepancy from the proposed formula can be seen, 
which is very likely due to stronger finite-size effect around this point.
}
\label{fig:xxz}
\end{center}
\end{figure}

We first focus on the $c=1$ Tomonaga-Luttinger liquid phase extending over a wide region
in $\Delta>-1$. 
The boson radius $R$ depends on $\Delta$ and $M$ 
(see Refs. \onlinecite{giamarchi} and \onlinecite{Bethe_XXZ3} for details).
When $M=0$, $R$ is related to $\Delta$ via a simple relation:
\begin{equation}
	R=\sqrt{2-\frac{2}{\pi}\arccos \Delta }~,~~~
    -1<\Delta \le 1.
	\label{eq:RDelta}
\end{equation}
When $M\neq 0$, $R$ can be determined numerically 
by solving the integral equations obtained 
from the Bethe-ansatz method.\cite{Bethe_XXZ1,Bethe_XXZ2,Bethe_XXZ3} 
We set $M$ at simple fractions $0,~1/5,~1/4,$ and $1/2$ 
so that we can examine the dependence on the system size $L$. 
As in the critical dimer models studied in the previous section, 
the entropy $S$ well obeys the scaling form $S \simeq \alpha L+ S_0 + b/L$. 
The subleading constant $S_0$ obtained by fitting the data 
is plotted as a function of $R$ in Fig.~\ref{fig:xxz}, 
which shows a remarkable agreement with 
\begin{equation}
 S_0 = \log R  -\frac{1}{2}.
\end{equation}

\begin{figure}
\begin{center}
\includegraphics[width=6cm,angle=-90]{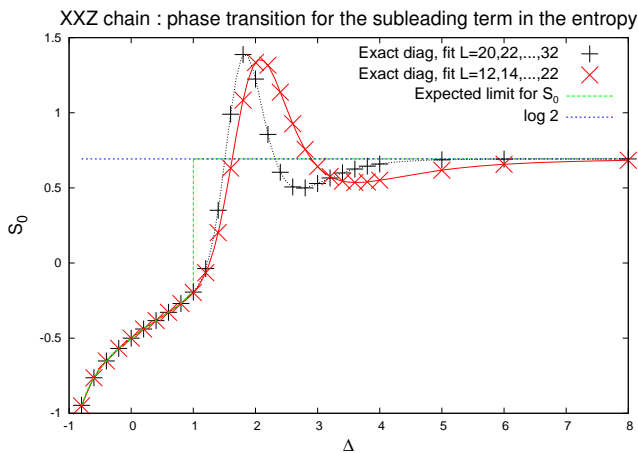}
\caption[1]{(color online) 
The constant part $S_0$ of the entropy and 
the phase transition in the XXZ chain at $M=0$. 
In the thermodynamical limit, we expect $S_0$ to be $S_0=\log R(\Delta)-1/2$ for
$-1<\Delta \leq 1$, 
and $S_0=\log 2$ for $\Delta>1$. 
}
\label{fig:xxz_transition}
\end{center}
\end{figure}

When increasing $\Delta$ at $M=0$, 
the system undergoes a Kosterlitz-Thouless transition at $\Delta=1$ 
from the critical phase to a massive N\'eel phase with doubly-degenerate ground
states. 
In a finite-size system, the double degeneracy in the N\'eel phase is slightly
split, 
and the ground state can be approximated by a macroscopic superposition of
ordered states. 
When $\Delta\to\infty$, such a state is given by 
\begin{equation}\label{eq:ud+du}
 |g\rangle = \frac1{\sqrt{2}} 
 \left(
  |\uparrow\downarrow\dots\rangle + |\downarrow\uparrow\dots\rangle
 \right). 
\end{equation}
This state gives $S=S_0=\log 2$. 
As in the discussion of Sec.~\ref{sec:transition}, 
one can expect that quantum fluctuations around the state \eqref{eq:ud+du}
occurring in $\Delta<\infty$ 
produce only extensive contributions 
and that the constant $S_0=\log 2$ is stable in the massive phase $\Delta>1$. 
Our numerical result for $S_0$ is presented in Fig.~\ref{fig:xxz_transition}.
The data show deviation from $\log 2$ when decreasing $\Delta$,  
but it is likely due to finite-size effects. 
The peak and dip seen in the figure move to the left 
as we use larger $L$'s for extracting $S_0$. 
In the thermodynamic limit, we expect a jump from 
$S_0=\log \sqrt{2} -1/2$ to $S_0=\log 2$ at the transition point $\Delta=1$.

Finally, we note that the XXZ chain with $\Delta=1/2$ and $h=0$ 
corresponds to the so-called ice model,\cite{Baxter82,Kasteleyn75}  
where all the configurations satisfying the ice rule (two-in and two-out around
every vertex) 
occur with equal probabilities. 
The RK state built from the ice model has been studied 
for the spin and fermionic models on the checkerboard
lattice.\cite{smp04,pbsf06} 
The result in this section shows that the half-cylinder entanglement entropy 
of this state has a subleading constant $S_0=\log \sqrt{4/3} -1/2$. 

\section{Ising chain in a transverse field}
\label{sec:ising_chain}

\newcommand{\xbasis}{{(x)}}
\newcommand{\zbasis}{{(z)}}

As an example showing a $c=1/2$ critical point, in this section, 
we study an Ising chain in a transverse field:
\begin{equation}
 \mathcal{H}=-\mu \sum_{j=0}^{L-1} \sigma^x_j \sigma^x_{j+1} - \sum_{j=0}^{L-1}
\sigma^z_j.
	\label{eq:Hictf}
\end{equation}
This model is related to two types of 2D classical models 
depending on which basis we work with.\cite{Kasteleyn75} 
In the $\sigma^x$ basis, we have a 2D Ising model:
\begin{equation}\label{eq:Ising2D}
 E = -\sum_{\la jj^\prime \ra} \sigma^x_j \sigma^x_{j^\prime},
\end{equation}
where $jj^\prime$ runs over all the nearest-neighbor pairs of sites 
on the square lattice. 
This model shows a low-temperature ordered phase and a high-temperature
paramagnetic phase. 
On the other hand, in the $\sigma^z$ basis, we have an eight-vertex model of
special type. 
The spins $\sigma^z_j$ are placed on the bonds of the square lattice 
and satisfy local constraints; the product of four spins around each site must
be even: 
\begin{equation}\label{eq:loc_const}
 \prod_{j\in \textrm{{\Large +}}} \sigma_j^z = +1. 
\end{equation}
Then the four spins can take $8$ states out of $2^4$ possibilities, 
hence the naming, eight-vertex. 
The energy is given by 
\begin{equation}\label{eq:8-vertex}
 E = - \sum_j \sigma^z_j.   
\end{equation}
It is useful to introduce a loop representation of the configurations.   
We regard the lowest-energy state ($\sigma^z_j=+1$ for all $j$) as the
``vacuum,''  
and place a loop element on every bond $j$ with $\sigma^z_j=-1$. 
Then only closed loops are formed because of the local constraints
\eqref{eq:loc_const}. 
Equation~\eqref{eq:8-vertex} means that the energy cost to generate loops is
proportional to their total length.  
At low temperatures, the system contains only small loops and is dominated by
the vacuum (``small-loop'' phase).
At high temperatures, the formations of large loops are allowed and the system
gets disordered (``large-loop'' phase). 
The correspondence among quantum and classical models is shown in
Table~\ref{table:Ising}, 
together with our results for the entropy which we present below.

Here we consider the Shannon entropies, $S^\zbasis$ and $S^\xbasis$, 
defined in the $\sigma^z$ and $\sigma^x$ bases respectively. 
These correspond to the half-cylinder entanglement entropies of the RK states 
built from the eight-vertex model and the 2D Ising model, respectively
(notice that for the latter, one needs to modify the model slightly 
in order to simplify the calculation, as presented in Sec.~\ref{sec:schmidt2}). 
The RK state constructed from the eight-vertex model \eqref{eq:8-vertex} in the
large-loop phase 
is particularly interesting because it possesses topological order. 
Such a state has been studied as the ground state of 
a quantum eight-vertex model\cite{prf07,aff04} (also known as an extended toric
code model\cite{cc07}).

\newcommand{\twolines}[2]{$\begin{matrix} \text{#1} \\ \text{#2} \end{matrix}$}
\begin{table*}
\caption{\label{table:Ising}
Correspondence between the Ising chain in a transverse field 
and related 2D classical models, 
and the results for the constant part of the entropy $S^\zbasis$ and
$S^\xbasis$. 
}
\begin{tabular}{c||c|c|c}
\hline\hline
 \twolines{Ising chain in a}{transverse field \eqref{eq:Hictf}}
 & disordered phase $\mu<1$  
 & $c=\frac12$ critical point $\mu=1$
 & ordered phase $\mu>1$ \\ 
\hline
 \twolines{Constant part}{of the entropy}
 &\twolines{$S^\zbasis_0=0$}{$S^\xbasis_0=0$}
 &\twolines{$S^\zbasis_0=-0.4387(1)$}{$S^\xbasis_0=S^\zbasis_0+\log 2=0.2544(1)$}
 &\twolines{$S^\zbasis_0=-\log 2$}{$S^\xbasis_0=+\log 2$}\\
\hline
 Eight-vertex model \eqref{eq:8-vertex}
 & \twolines{small-loop phase}{(low temperature)} 
 &
 & \twolines{large-loop phase}{(high temperature)}\\
\hline
 2D Ising model \eqref{eq:Ising2D}
 & \twolines{disordered phase}{(high temperature)}
 &
 & \twolines{ordered phase}{(low temperature)}\\
\hline\hline
\end{tabular}
\end{table*}

As is well known, the Hamiltonian \eqref{eq:Hictf} reduces to a fermionic
quadratic form 
using the Jordan-Wigner transformation. 
It can then be diagonalized using the Bogoliubov transformation 
(see Appendix \ref{sec:ictf}). 
The weight $|\la i|g\ra|^2$ of each spin configuration $|i\rangle$ in the
$\sigma^z$ basis 
can be obtained by calculating a Pfaffian, 
and $S^\zbasis(L)$ is computed numerically by summing over all the $2^L$
configurations. 

\begin{figure}
\begin{center}
\includegraphics[width=6cm,angle=-90]{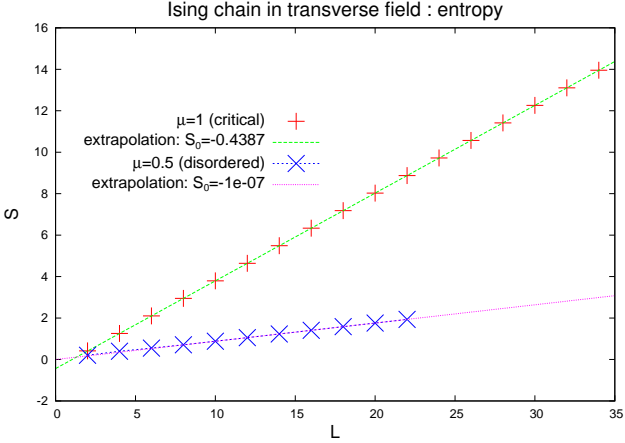}
\caption[1]{(color online) 
Entropy $S^\zbasis$ computed from the ground state of an Ising chain in a
transverse field 
up to $L=36$ spins. 
The data for $\mu=1$ (critical point) are well reproduced by 
$S^\zbasis\simeq \alpha L+S_0^\zbasis+\delta/L$ with $S_0^\zbasis\simeq-0.4387(1)$ 
(determined from a fit to the last three points $L=32,\,34,\,36$).
For $\mu=0.5$ in the disordered phase, the constant is very close to zero
(a fit to the three points $L=18,\,20,\,22$ gives $|S_0^\zbasis|\leq
10^{-6}$). 
}\label{fig:ictf}
\end{center}
\end{figure}

\begin{figure}
\begin{center}
\includegraphics[width=6cm,angle=-90]{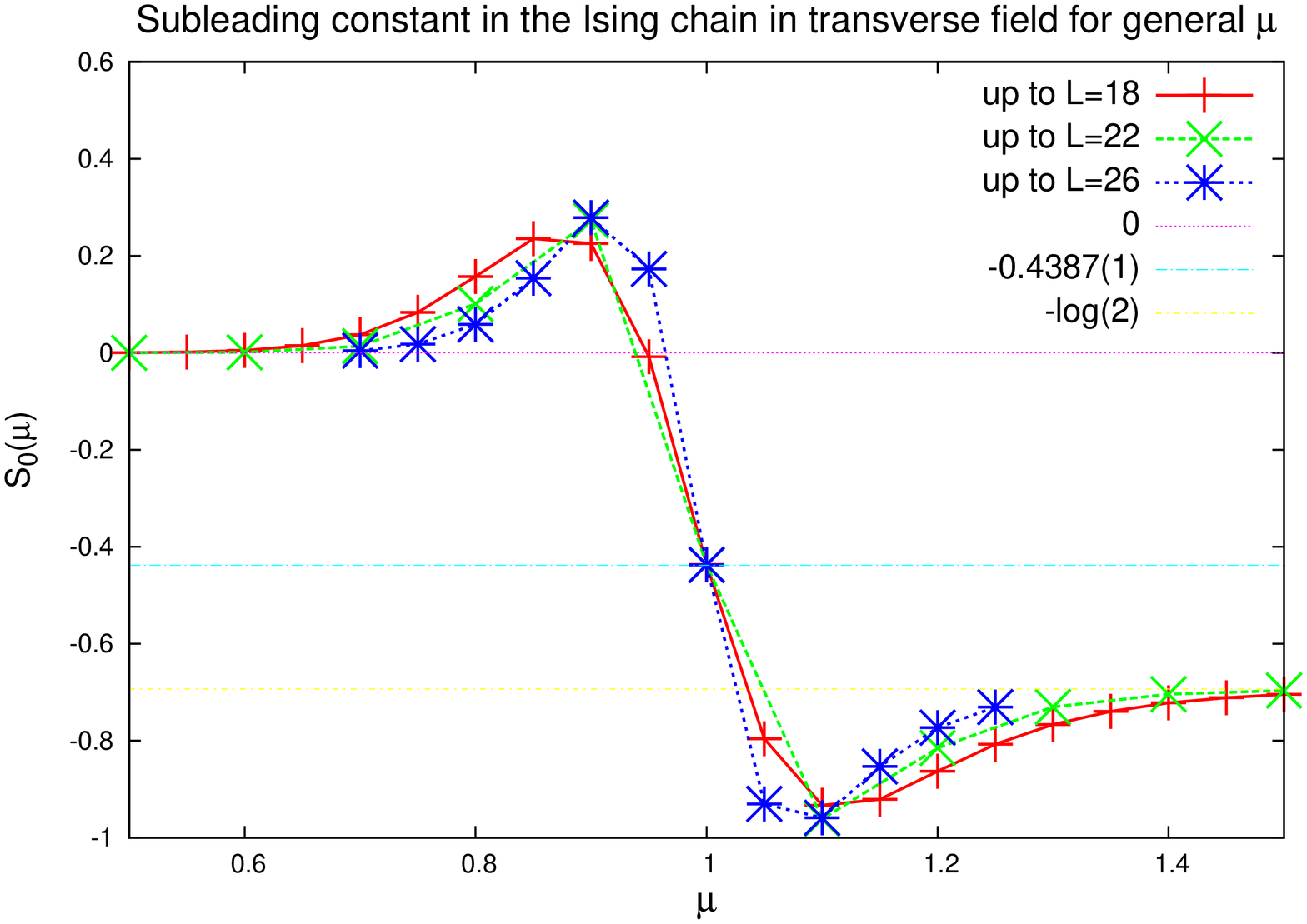}
\caption[1]{(color online) 
Subleading constant $S_0^\zbasis$ extracted from the entropy $S^\zbasis(L)$ 
in an Ising chain in a transverse field for different values of $\mu$. 
}\label{fig:Ising_nc}
\includegraphics[width=6cm,angle=-90]{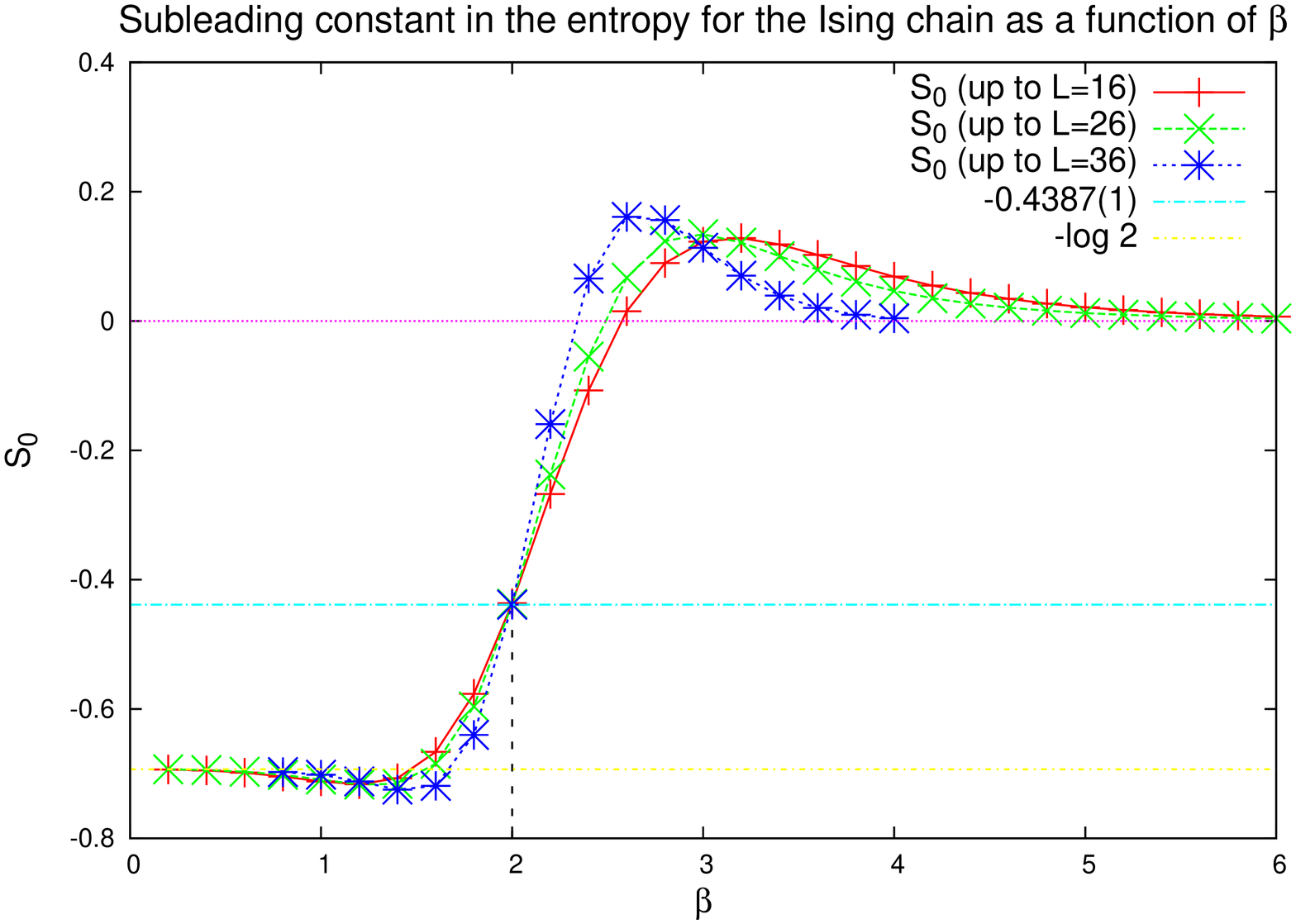}
\caption[2]{(color online) 
Subleading constant $S_0^\zbasis(\beta)$ extracted from the entropy
$S^\zbasis(L;\beta)$ 
in an Ising chain in a transverse field at the critical point $\mu=1$. 
The inverse temperature $\beta$ is introduced as explained in Sec.~\ref{sec:thermo_extension}. 
}\label{fig:Ising_beta}
\end{center}
\end{figure}

The scalings of $S^\zbasis(L)$ are shown in Fig.~\ref{fig:ictf}. 
We again observe nice agreement with a linear scaling 
$S^\zbasis (L) =\alpha L + S_0^\zbasis$ 
both in the critical and massive cases. 
At the critical point ($\mu=1$), we find the subleading constant in the entropy
to be
\begin{equation}
 S_0^\zbasis=-0.4387\pm 0.0001.
 \label{eq:result_ising}
\end{equation}
Figure~\ref{fig:Ising_nc} shows the constant part $S_0^\zbasis$ as a function of
the coupling constant $\mu$. 
Away from the critical point, the constant is stable at certain values: 
$S_0^\zbasis=0$ in the disordered phase ($\mu<1$) 
and $S_0^\zbasis=-\log 2$ in the ferromagnetic phase ($\mu>1$). 
Deviations from these values near the critical point 
are likely due to finite-size effects because they decay as we increase the
system size. 
These values can be understood by considering two limits. 
In the limit $\mu = 0$, the ground-state wave function is
$|g\ra = |\!\! \uparrow\uparrow\ldots \uparrow\rangle_z$, 
and the entropy $S^\zbasis$ is zero. 
In the limit $\mu \to \infty$, the wave function is
\begin{eqnarray}
 |g\ra 
 &=& \frac1{\sqrt{2}} ( 
       |\!\!   \uparrow   \uparrow \ldots   \uparrow \ra_x
      +|\!\! \downarrow \downarrow \ldots \downarrow \ra_x
     )  \\ \nonumber
 &=& \frac{1}{\sqrt{2^{L-1}}}\underset{\prod
\sigma_j^z=+1}{\displaystyle{\sum_{\sigma_1=\uparrow,\downarrow}\ldots} 
     \sum_{\sigma_L=\uparrow,\downarrow}} \left|\sigma_1\right\ra_z \otimes
\ldots \otimes \left|\sigma_L\right\ra_z,
\end{eqnarray}
where all the configurations have an even number of up spins in the $\sigma^z$
basis.  
Therefore, the entropy is 
\begin{equation}\label{eq:S_muinf}
 S^\zbasis (L) = (L-1)\log 2.
\end{equation}
As explained in Secs.~\ref{sec:transition} and \ref{sec:xxz_chain}, 
we expect that quantum fluctuations around these limits 
produce only extensive contributions and keep the subleading constants stable.

It is useful to interpret these results in terms of the eight-vertex RK state. 
When the temperature of the eight-vertex model is set to infinity, 
the corresponding RK state is an equal-amplitude superposition of all the loop
configurations. 
This is the ground state of Kitaev's toric code model.\cite{Kitaev03} 
Using the method of Ref.~\onlinecite{hiz05}, the half-cylinder entanglement
entropy 
is shown to scale exactly as Eq.~\eqref{eq:S_muinf}, 
and the constant part $S^\zbasis_0 = -\log 2$ can be interpreted as 
the topological entropy\cite{kp06,lw06} associated with $\mathbb{Z}_2$
topological order. 
Microscopically, $-\log 2$ comes from the fact 
that in any configuration, the loops cross an {\it even} number of time 
with the boundary separating the two half-cylinders. 
The result in Fig.~\ref{fig:Ising_nc} therefore demonstrates the stability of
this topological entropy 
for the half-cylinder geometry in the entire large-loop phase. 
The jump in the entropy around $\mu=1$ is interpreted as a breakdown of
topological order. 
We comment that the stability of topological entropy has also been studied on
the same wave function 
for the disk and annulus geometries in Refs.~\onlinecite{cc07} and
\onlinecite{prf07}.

We move on to the entropy $S^\xbasis$ in the $\sigma^x$ basis. 
It can be related to $S^\zbasis$ 
using the Kramers-Wannier duality transformation\cite{kw41}:
\begin{equation}
 \sigma^z_j \to \tilde{\sigma}^x_{j-1} \tilde{\sigma}^x_j, ~~
 \sigma^x_j \sigma^x_{j+1} \to \tilde{\sigma}^z_j,  
\end{equation}
by which ${\cal H}(\mu)$ is related to ${\cal H}(1/\mu)$. 
Here, $\tilde{\sigma}_j^z=-1$ is identified with a domain wall 
between $\sigma^x_j$ and $\sigma^x_{j+1}$. 
Taking into account the two-to-one correspondence 
between $\sigma^x$ and $\tilde{\sigma}^z$ configurations, 
one can show 
\begin{equation}
 S^\xbasis (\mu)=S^\zbasis (1/\mu)+\log 2.
\label{eq:sxsz}
\end{equation}
Hence we obtain the results summarized in Table~\ref{table:Ising}. 
Now we have a positive constant $S_0^\xbasis=\log 2$ in the ordered phase
$\mu>1$. 
This is a consequence of the macroscopic superposition of two ordered states, 
as discussed for the ordered phase of the XXZ chain in
Sec.~\ref{sec:xxz_chain}.

We have obtained two constants 
$S_0^\zbasis=-0.4387(1)$ and $S_0^\xbasis = S_0^\zbasis+\log 2 = 0.2544(1)$ 
at the critical point, depending on the choice of basis. 
We expect that these might be generic constants characterizing the $c=1/2$
criticality, 
although at present we do not have any analytical derivation of these numbers. 

We can also introduce a temperature $\beta^{-1}$ for the entropy $S^\zbasis$ 
as described in Sec.~\ref{sec:thermo_extension}. 
The constant part $S^\zbasis_0(\beta)$ extracted by fitting $S^\zbasis
(L;\beta)$ with a linear scaling 
changes rapidly around $\beta=2$ (see Fig.~\ref{fig:Ising_beta}). It seems reasonable to conjecture that
 $S_0^\zbasis(\beta)$ becomes a step function in the thermodynamic limit :
$S_0^\zbasis(\beta)=-\log 2$ 
for $\beta<2$ and $S_0^\zbasis(\beta)=0$ for $\beta>2$.
If confirmed, this result would suggest that increasing $\beta$ has a role similar to decreasing $\mu$,
 {\it i.e.} taking the system away from its critical point. In the $c=1$ case,
 $S_0(\beta)$ was a smooth function of $\beta$ (see Figs.~\ref{fig:hex_beta}, \ref{fig:phase_transition} and \ref{fig:xxz}).
 Therefore, it looks like introducing $\beta$ has a
 qualitatively different effect, depending on the nature of the critical theory.

Note that there are several directions in which to extend the Ising model. 
One possibility is to study the $q$-state Potts model or the RSOS models along
the same lines. 
Another one is to view the Ising model as a special case ($n=1$) of the dilute
$\mathcal{O}(n)$ loop model. 
In the loop model case, the $p(\mu_i)$ are the probabilities 
that the equator of the sphere is run across by loops at positions $\mu_i$. 
In that case, we would find a universal curve $S_0(n)$ extending
Eq.~\eqref{eq:result_ising}.

\section{Scaling of the largest probability}
\label{sec:scaling_pmax}

In this section, we study the scaling of the largest probability 
\begin{equation}
 p_0  := \mathop{\max}_i p_i=| \la i_0 | g \ra |^2, 
\end{equation}
i.e., the weight of the most probable configuration $i_0$ in the 1D wave
function $|g\ra$. 
In terms of a 2D RK state, 
this corresponds to the largest eigenvalue of the reduced density matrix
$\rho_A$ of a half cylinder.
Very similarly to the entropy $S$, we find that $-\log p_0$ exhibits  
a linear scaling with $L$ followed by a subleading universal constant: 
\begin{equation}\label{eq:C}
 -\log p_0=\tilde{\alpha} L + \gamma+o(1). 
\end{equation}
Below we evaluate $\gamma$ in some critical systems. 

\subsection{$c=1$ critical systems}

\begin{figure}
\begin{center}
\includegraphics[angle=-90,width=8cm]{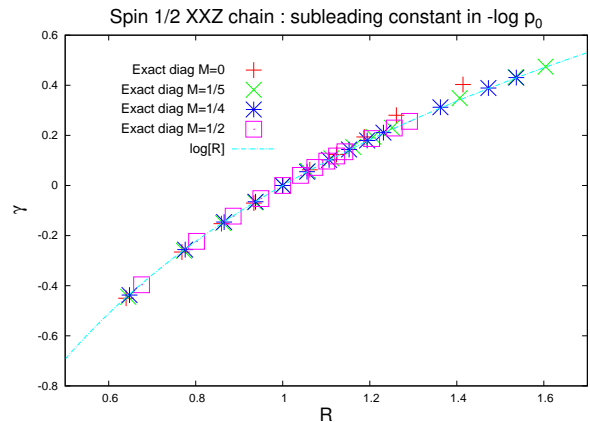}
\caption[1]{(color online)
The subleading constant $\gamma$ in the scaling of $-\log p_0$ [see
Eq.~\eqref{eq:C}] 
extracted from the critical ground state of the XXZ chain in a magnetic field. 
The same setting as Fig.~\ref{fig:xxz} is taken.

}
\label{fig:Hsu_XXZ}
\end{center}
\end{figure}

\begin{figure}
 \begin{center}
  \includegraphics[angle=-90,width=8cm]{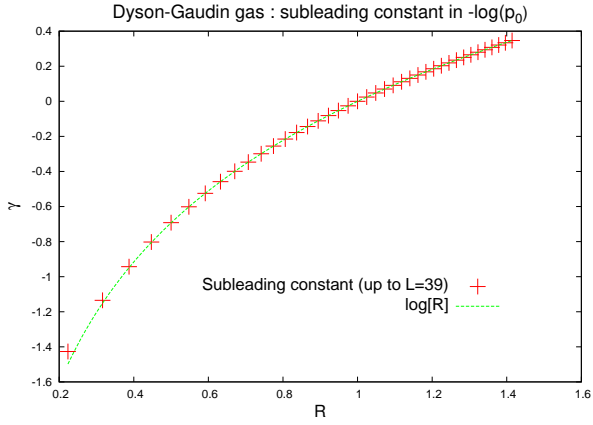}
\caption[1]{(color online) 
The subleading constant $\gamma$ in the scaling of $-\log p_0(\beta)$, 
where $p_0 (\beta)$ is the largest probability in the Dyson-Gaudin gas 
with density $\rho=1/3$. 
The data are consistent with Eq.~\eqref{eq:gamma_R}. 
For some special values of $R$ (see the text and
Appendix~\ref{sec:gaudin_annexe}), 
Gaudin's formula \eqref{01-gaudin} can be used to show Eq.~\eqref{eq:gamma_R}
exactly.
}
\label{fig:Hexa_beta_logz_bis}
 \end{center}
\end{figure}

We first consider the XXZ chain in a magnetic field, Eq.~\eqref{eq:XXZ}, 
in the critical phase.
We find that the largest probability 
$p_0 = |\la i_0|g\ra|^2$
is attained by crystal states. 
For example, 
$|i_0\ra = |\!\!\uparrow\downarrow\uparrow\downarrow\!\dots\ra$ and 
$|\!\!\downarrow\uparrow\downarrow\uparrow\!\dots\ra$ for $M=0$, 
and 
$|i_0\ra =
|\!\!\uparrow\uparrow\uparrow\downarrow\uparrow\uparrow\uparrow\downarrow\!\dots
\ra$, etc. 
for $M=1/2$, 
independent of $\Delta~(>-1)$. 
The constant $\gamma$ is extracted by fitting finite-size data with
Eq.~\eqref{eq:C}.
As shown in Fig.~\ref{fig:Hsu_XXZ}, 
we observe a simple relation:
\begin{equation}\label{eq:gamma_R}
 \gamma=\log R.
\end{equation}
The same result can be shown exactly for the largest probability 
$p_0(\beta):=\max p_i(\beta)$ in the Dyson-Gaudin gas 
when $\beta=2R^2=2\lambda$ with $\lambda \in \mathbb{N}$ and
$\lambda<\frac{L}{n-1}$; 
see Appendix \ref{sec:gaudin_annexe}. 
For general $\beta$, Eq.~\eqref{eq:gamma_R} can be numerically demonstrated 
as shown in Fig.~\ref{fig:Hexa_beta_logz_bis}.

In order to understand the connection between $S$ and $-\log p_0$, 
it is useful to introduce the R\'enyi entropy:
\begin{equation}\label{eq:Renyi}
 S^{(N)} := \frac{-1}{N-1} \log \left( \sum_i p_i^N \right) ,
\end{equation}
where $N$ is a real number.
Then, $S$ and $-\log p_0$ correspond to the limits  
$N\to 1$ and $\infty$, respectively. 
Now we assume that the probability distribution is given by 
the Boltzmann weights $p_i (\beta)=p_i^{\beta/2} / Z(\beta)$ of the Dyson-Gaudin
gas 
in the critical phase $R=\sqrt{\beta/2} \le d$. 
The R\'enyi entropy \eqref{eq:Renyi} can then be expressed as 
\begin{equation}
 S^{(N)} (\beta) = \frac{-1}{N-1} \log \frac{Z( N\beta )}{[Z(\beta)]^N}. 
\end{equation}
Recalling Eq.~\eqref{eq:subleading_c} for the non-extensive part $C$ of
$Z(\beta)$,  
the subleading constant contribution to $S^{(N)}$ is given by 
\begin{equation}\label{eq:renyi_const}
 \left\{
  \begin{array}{ccc}
   \log R - \frac{1}{2(N-1)}\log N, & & \sqrt{N}R<R_c=d, \\ 
   \frac{N}{N-1}\log R - \frac1{N-1} \log d, & & \sqrt{N}R> R_c=d.
  \end{array}
 \right.
\end{equation}
Both expressions give  $\gamma=\log R$ in the limit $N\to \infty$. 
On the other hand, the former expression is consistent with $S_0=\log R-1/2$ 
in the limit $N\to 1$.

An alternative strategy to derive Eq.~\eqref{eq:gamma_R} is 
to adopt the 2D viewpoint of Sec.~\ref{sec:RK_states}, 
where the probability $p_0$ was related to classical partition functions: 
\begin{equation}\label{eq:p0_cl}
 -\log p_0 = -\log \frac{ \mathcal{Z}_{i_0}^A \mathcal{Z}_{i_0}^B }{ \mathcal{Z}
}.  
\end{equation}
Here, $\mathcal{Z}_{i_0}^A$ and $\mathcal{Z}_{i_0}^B$ are partition functions on
$A$ and $B$  
with spins fixed in a state $i_0$ at their common boundary. 
Now we move on to the continuum limit described by the action
\eqref{eq:FreeFieldAction}. 
Recalling that $i_0$ is given by a crystal state, 
the above boundary condition corresponds to locking the field $\phi$ 
at a certain constant at the boundary (Dirichlet boundary condition). 
\footnote{
One can recall that at a certain value of $R$, 
the sine potential for $\phi$ becomes relevant 
and $\phi$ is locked at a constant value, leading to a crystallization.
Conversely, a crystal state can be regarded as a $\phi$-locked state. 
}
Hence, we obtain 
\begin{equation}\label{eq:p0_Dirichlet}
 -\log p_0 = -\log \frac{ \mathcal{Z}_D^A \mathcal{Z}_D^B }{ \mathcal{Z} },  
\end{equation}
where $D$ stands for Dirichlet. 
This expression has been evaluated 
by Hsu {\it et al.}\cite{hsu08} and by Campos Venuti {\it et al.}\cite{cvsz09} 
using boundary CFT. 
Their results for the non-extensive part are consistent with
Eq.~\eqref{eq:gamma_R}. 
\footnote{
Note that for the dimer models on the square and hexagonal lattices, 
the radius of $\phi$ is given by $R=1$, 
not by $R=2$ used by Hsu {\it et al.}\cite{hsu08} 
}

In fact, Hsu {\it et al.}\cite{hsu08} proposed the right hand side of
Eq.~\eqref{eq:p0_Dirichlet} 
as the expression of the entanglement entropy $S^{\rm VN}$ of a half cylinder. 
Their argument was based on a replica trick; 
the $N$-th moment ${\cal M}^{(N)}:={\rm Tr}~\rho_A^N = \sum_i p_i^N$ of the
reduced density matrix 
was evaluated for integer $N\ge 2$ 
and then an analytic continuation $N\to 1$ was taken.
According to their evaluation, the R\'enyi entropy $S^{(N)}=\frac{-1}{N-1}\log
{\cal M}^{(N)}$ 
is expressed by the r.h.s. of Eq.~\eqref{eq:p0_Dirichlet} for any integer
$N\ge2$, 
leading to an {\it $N$-independent} subleading constant $\log R$. 
On the other hand, we have obtained {\it $N$-dependent} constant
\eqref{eq:renyi_const}. 
The two results for the subleading constant agree only in the limit
$N\to\infty$.  
We infer that this discrepancy comes from a difficulty 
in specifying boundary conditions in the argument of Hsu {\it et
al.}\cite{hsu08} 
In their argument, they took linear combinations of plural compactified fields, 
which could make the compactification conditions ambiguous. 
A more careful treatment of the compactification conditions 
and a derivation of Eq.~\eqref{eq:renyi_const} in boundary CFT 
are left as important open issues.

\subsection{$c=1/2$ critical system and beyond}

\begin{figure}
 \begin{center}
  \includegraphics[angle=-90,width=8cm]{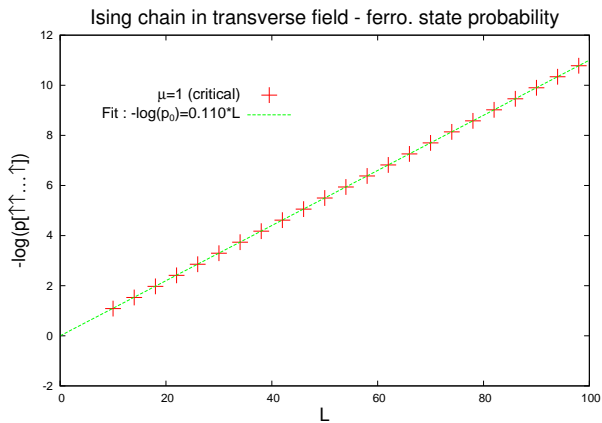}
 \end{center}
\caption[1]{(color online) 
Scaling of $-\log p_0^\zbasis$ in the Ising model in a transverse field 
at the critical point, 
calculated in the $\sigma^z$ basis. 
The subleading constant $\gamma^\zbasis$ is very close to zero:  
a fit to the last three points $L=96,\,98,\,100$ gives $|\gamma|\leq 10^{-6}$.
}\label{fig:ising_proba}
\end{figure}

We next consider the Ising chain in a transverse field, Eq.~\eqref{eq:Hictf}, 
at the $c=1/2$ critical point $\mu=1$. 
Figure~\ref{fig:ising_proba} shows the scaling of $-\log p_0^\zbasis$ in the $\sigma^z$
basis. Here, the largest probability is attained by the ferromagnetic configuration 
$|i_0\ra = |\!\!\uparrow\uparrow\uparrow\!\!\ldots\rangle_z$. 
In this case we have the following exact formula (see Sec.~\ref{sec:ictf_pmax}):
\begin{equation}
 p_0^{(z)}=\prod_{j=1}^{L/2} \cos^2\left(\frac{(2j-1)\pi}{4L}\right)
\end{equation}
An Euler-Maclaurin expansion shows that the subleading constant is $\gamma^{(z)}=0$. 
The data in Figure~\ref{fig:ising_proba} is consistant with this result and gives $\gamma^{(z)}<10^{-6}$.
 It can also be shown that $\gamma^{(z)}$ remains zero away from the critical point.
Moving to the $\sigma^x$ basis using the same argument leading to Eq.~\eqref{eq:sxsz}, 
we obtain $\gamma^{(x)}=\log 2$.

In a 2D viewpoint, $p_0$ is related to a ratio of partition functions, 
as shown in Eq.~\eqref{eq:p0_cl}. 
Since the boundary configurations, $a_h$ and $b_h$, 
at the upper and lower edges of the cylinder were arbitrary, 
we can glue these edges by identifying $a_h$ and $b_h$ and integrating them out. 
Algebraically, we consider 
\begin{equation}\label{eq:p0_ZZ}
 p_0 
 = \frac{\la i_0| {\cal T}^{2h} |i_0 \ra}{{\rm Tr}~{\cal T}^{2h}}
 = \frac{{\cal Z}_{i_0i_0}(L,2h)}{{\cal Z}_P(L,2h)} ~~~
 (h\to\infty).
\end{equation}
Here, the numerator ${\cal Z}_{i_0i_0}$ is the partition function 
of a long cylinder with boundary configurations fixed in the same state $i_0$ at
both edges. 
The denominator ${\cal Z}_P$ is the partition function of a torus. 

A similar quantity has been considered in a rational CFT context. 
Therein, the fixed boundary conditions imposed in the numerator of
Eq.~\eqref{eq:p0_ZZ} 
are replaced by conformally invariant ones $a$.
The associated probability $p_a$ can be evaluated as explained in
Ref.~\onlinecite{zuber}. 
In the notations of Ref.~\onlinecite{zuber}, the result is:
\begin{eqnarray}
 p_a=\frac{(\psi_a^1)^2}{S_1^1}.
  \label {dimension}
\end{eqnarray}
where $\psi_a^i$ are certain structure constants characteristic of the model, 
and $S_1^1$ is the identity matrix element of the S matrix implementing the
modular transformation. 
In the simplest case of an $A_m$ $SU(2)_k$ model, 
the boundary fields correspond to the vertices of the $A_m$ Dynkin diagram. 
Let $\{d_a\}$ be the Perron-Frobenius eigenvector of the $A_m$ incidence
matrix, 
normalized so that $d_1=1$. 
Then, each $d_a$ is the so called quantum dimension of the state $a$, 
and $d=\sqrt {\sum_a d_a^2}$ is called the total quantum dimension. 
In this case, $\psi_a=d_a/d$ and $S_1=1/d$, hence
\begin{eqnarray}
 p_a=\frac{d_a^2}{d}.
  \label {dimension1}
\end{eqnarray}
We study the case where the Dynkin diagram is $A_3$ and the possible states are $+,~{\rm free},~-$  
with quantum dimensions $1,~\sqrt{2},~1$, respectively. 
The probabilities are, in the $\sigma^x$ basis
\begin{eqnarray}
 p_+=p_{-}=\frac{1}{2},\ p_{\rm free}=1.
 \label {dimension1bis}
\end{eqnarray}

$A_3$ also describes the Ising model and this enables us to confirm the numerical results at the beginning of this
subsection.
The ferromagnetic state in the $\sigma^z$ basis,
$|\uparrow\uparrow\uparrow\ldots\uparrow\rangle_z$, 
may be regarded as a paramagnetic state in the $\sigma^x$ basis or the ``free''
state in CFT. 
Equation~\eqref{dimension1bis} then gives 
\begin{equation}
 \gamma^{(z)}=-\log p_{\rm free}=0, 
\end{equation} 
in agreement with our numerical result. 
The largest probability $p_0$ in the $\sigma^x$ basis 
may be regarded as the probability of the ``+'' state in CFT, hence
\begin{equation}
 \gamma^{(x)}=-\log p_+=\log 2 
\end{equation}
in agreement with our calculation. \\
These results can be extended to $A_m$ RSOS models with central charge
$c<1$. The configuration $p_0$ with highest probability is in this case
\begin{equation}
 p_0=2\sqrt{\frac{2}{m(m+1)}}\sin \left(\frac{\pi}{m+1}\right)\sin \left(\frac{\pi}{m}\right).
\end{equation}

\section{Summary and conclusions}

The starting point of this study was to introduce the Shannon entropy of a 1D ground state wave function, 
which measures quantum fluctuations occurring in a given basis. 

Like other entanglement measures, we have seen that the scaling behavior of this entropy is essentially 
controlled by the long-distance correlations. 
Using a transfer matrix approach, we showed that this entropy 
can also be interpreted as the entanglement entropy of a half cylinder for a suitably chosen 2D RK state. 
This correspondence allowed us 
to study the entanglement entropy of 2D wave functions 
using simpler 1D systems, 
without the need to trace explicitly over the degrees of freedom sitting outside of the subsystem 
(a formidable task in 2D). 

To unveil the generic scaling properties of the Shannon entropy of 1D states 
(equal to the entanglement entropy of 2D states), 
we have studied several 1D quantum systems: 
(i) a discretized version of the Dyson gas/Calogero-Sutherland ground state wave function  
(relevant to 2D dimer RK states) in Sec.~\ref{sec:RKtoGaudin}, 
(ii) the spin-$\frac{1}{2}$ XXZ chain 
(relevant to six-vertex RK states) in Sec.~\ref{sec:xxz_chain}
and (iii) the Ising chain in transverse field 
(relevant to 2D Ising RK states and 2D eight-vertex RK states) in Sec.~\ref{sec:ising_chain}.

In both critical and massive systems, 
we found that this entropy is composed of an extensive part and a subleading constant $S_0$. 
There is no logarithmic contribution as anticipated before for half-cylinder geometry.\cite{fm06,hsu08}
For Tomonaga-Luttinger liquids (cases (i) and (ii) above),
described by a compactified boson with radius $R$, we showed 
numerically and analytically 
that $S_0=\log R - \frac{1}{2}$
(a result which {\it differs} from the recent prediction by Hsu {\it et al.}\cite{hsu08}).
Going back to the 2D entanglement entropy interpretation of this result, it implies that the usual RK states 
for dimers on the hexagonal or square lattice (with $R=1$) have $S_0=-\frac{1}{2}$. 
At present, we do not have a derivation for the value $S_0^{(z)}=-0.4387$ (or
$S_0^{(x)}=-0.4387+\log 2$ depending
on the choice of the basis) found numerically for the Ising chain in
transverse field at its $c=1/2$ critical point.

We introduced a temperature $\beta^{-1}$ 
to extend the Shannon and entanglement entropies in Sec.~\ref{sec:thermo_extension}. 
This has different consequences depending on the nature of the criticality. 
For TLLs, 
changing $\beta$ gives a natural way to tune the boson radius $R$ 
while retaining the $c=1$ criticality. 
This allowed us to ``deform'' a dimer RK problem (or a free fermion problem) 
and to derive the $R$-dependence of the entropy constant $S_0$ in Sec.~\ref{sec:RKtoGaudin}. 
When $\beta$ reaches a critical value, 
the system undergoes a phase transition to a crystal state, 
where the entropy constant takes a stable value $S_0=\log d$ 
with $d$ being the degeneracy of the ground states. 
Hence, this transition can be detected through $S_0$. 
In contrast, in the critical Ising chain with $c=1/2$, 
the entropy constant $S_0$ shows an abrupt change around $\beta=2$. 
More generally, the $\beta$-dependence might offer 
a valuable fingerprint for clarifying the nature of the undeformed case $\beta=2$.

We also considered the scaling properties of $p_0$, 
the probability of the  most likely configuration in a critical 1D state
(Sec.~\ref{sec:scaling_pmax}).
This quantity, which corresponds to the largest eigenvalue of the
reduced density matrix in the 2D point of view, 
also contains a universal constant contribution $\gamma$ in critical states. 
We found numerically and analytically 
that $\gamma=\log R$ for TLL states 
and $\gamma^{(z)}=0$ and $\gamma^{(x)}=\log 2$ for the critical Ising chain.
These are related to the probabilities associated with conformally invariant boundary conditions.  

The universal $R$-dependence of $S_0$ and $\gamma$ found in the present work 
can be used as a new simple way to determine the boson radius $R$ in a TLL 
through a ground-state structure. 
Similar universal $R$-dependence was also found 
in the mutual information (double-interval entanglement entropy)
studied in Refs.~\onlinecite{fps09} and \onlinecite{cct09}. 
Notice that the mutual information is invariant under the transformation $R \rightarrow 2/R$ 
while the present quantities are not. 
This difference is related to the origins of the $R$-dependence: 
it comes from certain boundary effects in the present case, 
while it comes from the special topologies of the Riemann surfaces 
in the case of the mutual information.\cite{fps09,cct09}

In fact, the transfer matrix approach 
gives access to {\it all} the eigenvalues $\{p_i\}$ of 
reduced density matrix $\rho_A$ of a half cylinder. 
(as a first application, the gap $p_1/p_0$ is computed in Appendix \ref{sec:gaudin_annexe} 
for dimers on the hexagonal lattice). 
The present approach therefore provides a convenient tool 
to study the properties of the ``entanglement spectrum''\cite{lh08}
in a 2D quantum state.

{\it Acknowledgments --- } 
We wish to thank J\'er\^ome Dubail, Michel Gaudin, Benjamin Hsu, Bernard
Nienhuis St\'ephane Nonenmacher, and Keisuke Totsuka for fruitful discussions.

\appendix{}

\section{Free fermions and Wick's theorem}
\label{sec:wick}
The probabilities of Eq.~\eqref{eq:pi} 
involve  quantities such as $\la 0|a_1 a_2\ldots a_{2n}|0\ra$, where the $a_j$
are linear combinations of fermion creation and annihilation operators. Wick's
theorem then gives:
\begin{eqnarray}\nonumber
\la 0|a_1\ldots a_{2n}|0\ra&=&\la a_1\ldots a_{2n}\ra\\\nonumber
&=&\!\!\sum_{\begin{array}{c}i_1<\ldots <i_n\\\forall
k,i_k<j_k\end{array}}\!\!\!\!\epsilon(\sigma)\la a_{i_1}a_{j_1}\ra \ldots \la
a_{i_{n}}a_{j_{n}}\ra\\ \label{eq:wick}
&=&\textrm{Pf }A,
\end{eqnarray}
where $\epsilon(\sigma)$ is the signature of the permutation which transforms
$\{1,2,\ldots,2n\}$ into $\{i_1,j_1,i_2,j_2,\ldots i_n,j_n\}$. $\textrm{Pf}$
denotes the Pfaffian. $A$ is an antisymmetric $2n \times 2n$ matrix given by :
\begin{equation}
 A=\left[\begin{array}{cccc}0&\la a_1a_2\ra&\ldots&\la a_1 a_{2n} \ra\\-\la
a_1a_2\ra &0 &\ldots &\la a_2 a_{2n}\ra\\\vdots&\vdots&\ddots&\vdots\\-\la
a_1a_{2n}\ra &-\la a_2a_{2n}\ra &\ldots &0\end{array}\right].
\end{equation}
The two following properties of Pfaffians are useful :
\begin{eqnarray}
 \left(\textrm{Pf }A\right)^2&=&\det A\\ \label{E:pf}
\textrm{Pf }\left[\begin{array}{cc}0 &B\\-B^T&0\end{array}\right]&=&\pm \det B
\end{eqnarray}
and allow fast numerical calculations using determinant routines.

\section{Transfer matrix for the classical dimer model on the hexagonal
lattice}
\label{sec:hexa_tm}

\begin{figure}
\begin{center}
\includegraphics[width=8cm]{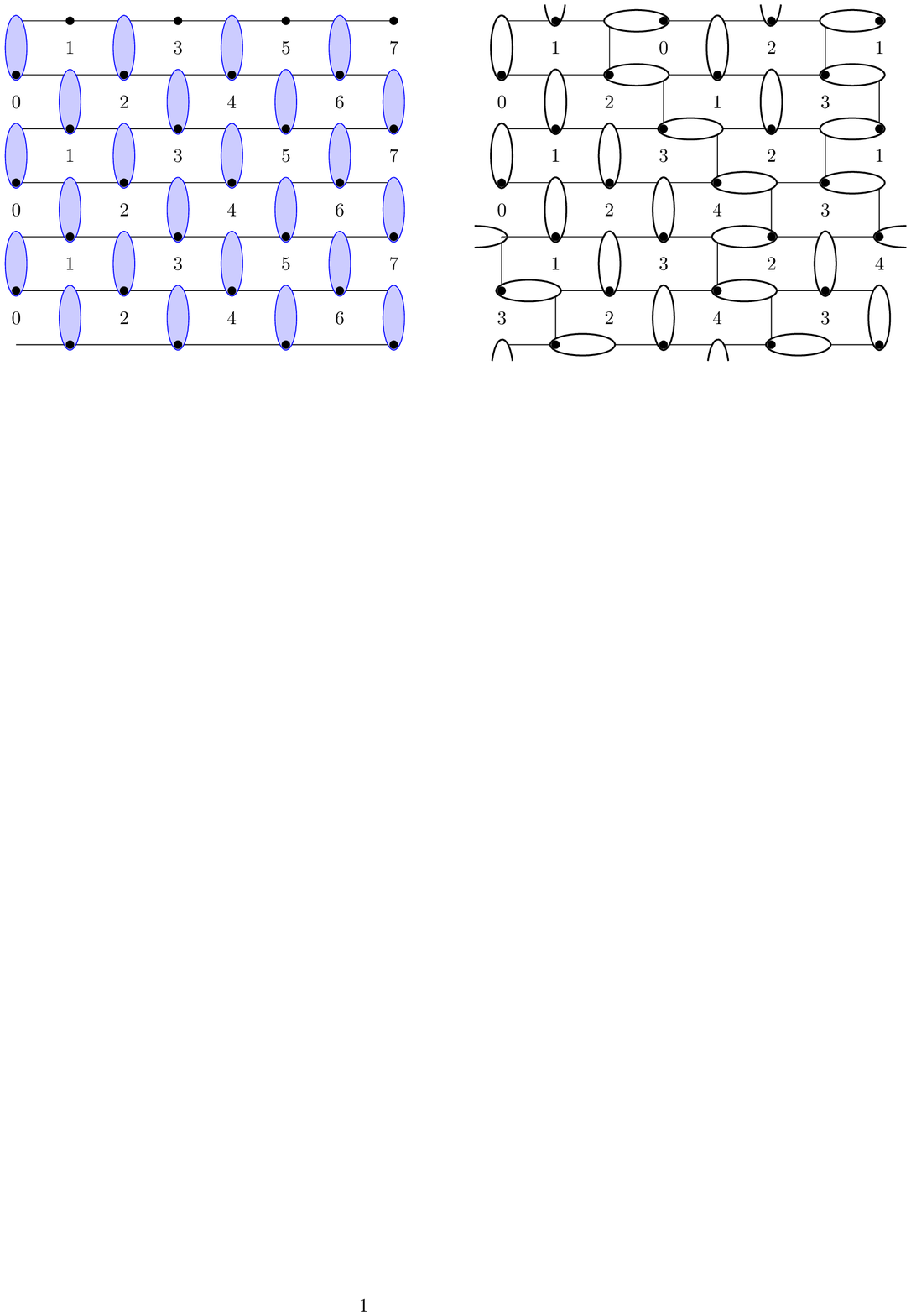}

\end{center}
\begin{center}
 \includegraphics[width=4cm]{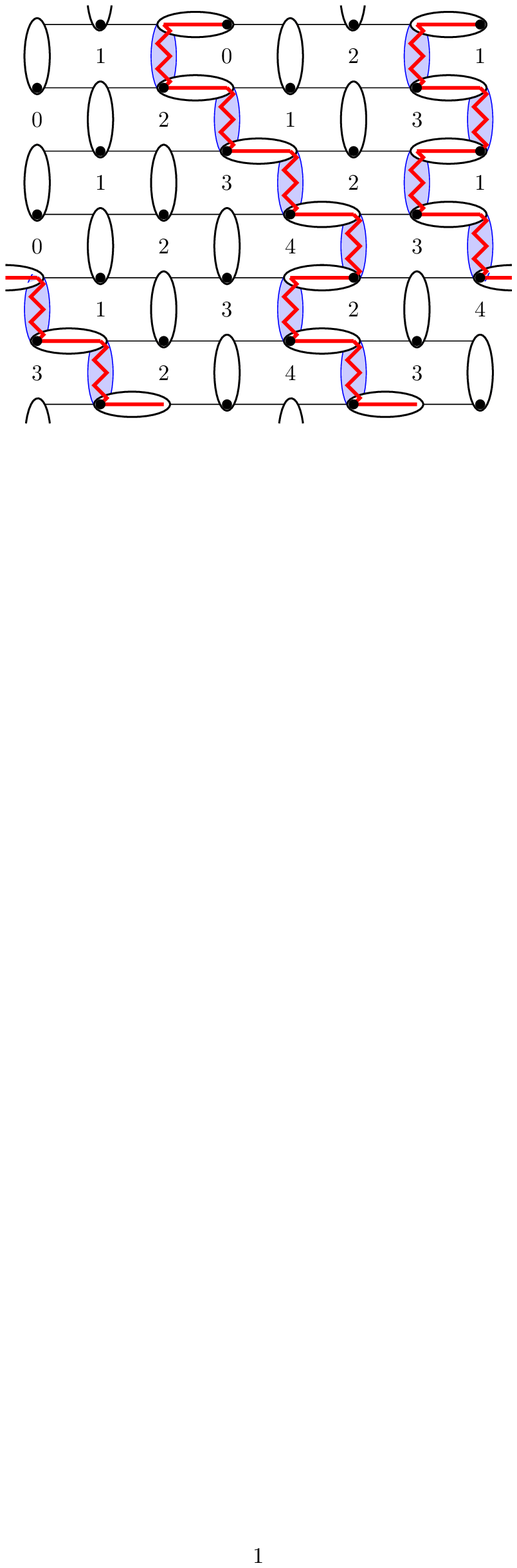}
\end{center}
\caption[1]{(color online)  Upper left : reference configuration. Upper right :
real configuration. Below : Transition graph. Reference dimers are in blue. The fermions are
living on the vertical edges of the lattice and are symbolized by red zigzag
lines.
The integers attached to each plaquette of the lattice
form a height configuration associated to the dimer covering.
When coarse-grained, these
microscopic heights become the free field which describe the long-distance properties of the system.\cite{blote,henley}
The heights can be constructed by fixing $h=0$ at some origin and then moving from plaquette to plaquette by turning clockwise around the sites of the even sublattice (marked with a black dot).
The rule is the following: the height picks a contribution equal to +2 when crossing a dimer,
and -1 otherwise. Since there is exactly one dimer touching each site, the height difference between
two points does not depends on the chosen path on a simply connected domain.
With periodic boundary conditions, the height is not single-valued.
For example, when winding horizontally around the system, the height picks a contribution $W_x$
(also called winding number)
equal to twice the number of vertical dimers crossed, minus the number of empty bonds.
Inserting a fermionic world line going upward
shifts the height by $-3$ by going from the left to the right, and thus changes $W_x \to W_x-3$.
It is simple to check that the configurations with a fermion density equal to $\frac{2}{3}$ have $W_x=0$ and an average ``slope'' equal
to zero.

}\label{fig:Hexagone}
\end{figure}
\subsection{Transfer matrix as free fermions}
Here we consider a hexagonal lattice with periodic boundary conditions and an
even number of columns $L$. The mapping onto free fermions is as follows (see
Fig.~\ref{fig:Hexagone}):

We choose a convenient dimer configuration  which we call a reference
configuration. Any other dimer configuration (real configuration) will be
compared to the reference by superposition of the two. We define the particle
locations as the vertical edges that are {\it not} occupied by a ``real'' dimer
(only a reference one). Particles can jump from a vertical edge to another only
if a real horizontal dimer connects the two. This mapping has several
interesting properties :
\begin{itemize}
 \item The dimer configuration is totally determined by the trajectories of the
particles.
 \item Two particles cannot go to the same edge. Therefore, they obey a
fermionic exclusion rule. This encodes the dimer hardcore constraint.
  \item The number of fermions is conserved, so that the TM is block diagonal,
each block corresponding to a fixed number of fermions. It should be remarked
that this property would not hold on non-bipartite lattices (such as the
triangular).
\end{itemize}

\subsection{Fermionic representation and periodic boundary conditions}
A state of a row is determined by the number $n$ of fermions and their positions
$0\leq \alpha_1<,\ldots<\alpha_n\leq L-1$. We can choose to represent such a
state using second-quantized fermions creation operators :
\begin{equation}
 |i\ra=|\alpha_1,\ldots \alpha_n\ra=c_{\alpha_1}^\dag \ldots c_{\alpha_n}^\dag
|0\ra.
\label{eq:order}
\end{equation}
As we want to use the translational invariance, we have to set
\begin{eqnarray}
 |\alpha_1,\ldots,\alpha_{n-1},L\ra=|0,\alpha_1,\ldots,\alpha_{n-1}\ra.
\end{eqnarray}
Therefore, to keep the order of Eq.~\eqref{eq:order} :
\begin{equation}
c_{L}^\dag =(-1)^{\hat{n}-1}c_{0}^\dag \quad,\quad \hat{n}=\sum_{j=0}^{L-1}
c_j^\dag c_j.
\end{equation}
In the following, we will also need the $L$ fermion operators in Fourier space
:
\begin{equation}
 c_k^\dag=\frac{1}{\sqrt{L}}\sum_{j=0}^{L-1} e^{-ikj}c_j^\dag.
\end{equation}
They satisfy $\{c_k,c_{k'}^\dag\}=\delta _{kk'}$, provided
$e^{ikL}=(-1)^{\hat{n}-1}$. The set of wave-vectors is given by 
\begin{eqnarray}\nonumber
k \in \bigg\{-\pi+\frac{\pi}L+\frac{2\pi l}L\, \bigg| \;l=0,\ldots, L-1\bigg\}&,&\hat{n}\textrm{ even}
\\\nonumber
k \in \bigg\{-\pi+\frac{2\pi l}L\, \bigg| \;l=0,\ldots, L-1\bigg\}&,&\hat{n}\textrm{ odd}.
\end{eqnarray}

\subsection{Diagonalization of the Transfer matrix}
Each fermion can go to the left or to the right with equal amplitude. We number
vertical edges in such a way that a fermion located on $j$ can go to $j$ or
$j+1$. $\mathcal{T}$ satisfies 
\begin{eqnarray}
\mathcal{T}|0\ra&=&|0\ra\\
 \mathcal{T}c_j^\dag \mathcal{T}^{-1}&=&c_j^\dag +c_{j+1}^\dag.
\end{eqnarray}
So that
\begin{equation}
 \mathcal{T}c_k^\dag \mathcal{T}^{-1}=\lambda(k) c_k^\dag\quad,\quad
\lambda(k)=1+e^{ik}.
\end{equation}
$c_k^\dag|0\ra$ is then eigenvector of $\mathcal{T}$ with eigenvalue
$\lambda(k)$. In a similar manner :
\begin{equation}
 \mathcal{T}c_{k_1}^\dag c_{k_2}^\dag\ldots
c_{k_n}^\dag\mathcal{T}^{-1}=\lambda(k_1)\ldots \lambda(k_n)c_{k_1}^\dag \ldots
c_{k_n}^\dag.
\end{equation}
Provided all the wave vectors are different, $c_{k_1}^\dag \ldots c_{k_n}^\dag
|0\ra$ is an eigenvector with eigenvalue $\lambda(k_1)\ldots\lambda(k_n)$. The
transfer matrix can also be expressed explicitly :
\begin{equation}
 \mathcal{T}=\prod_k \left(1+e^{ik}c_k^\dag c_k\right).
\end{equation}

\subsection{Largest eigenvalue and dominant eigenvector}
\begin{figure}
\begin{center}
\includegraphics[angle=270,width=8cm]{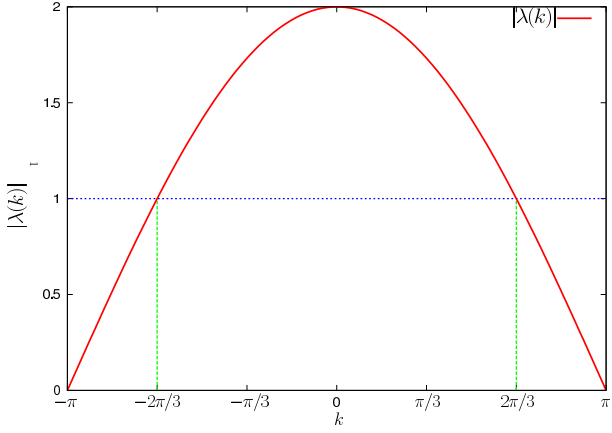}
\end{center}
\caption[1]{(color online) One-particle eigenvalue as a function of
$k$}\label{fig:Disp_hex}
\end{figure}
Since  $|\lambda(k)|\geq1$ for every $k \in
[-2\pi/3,2\pi/3]$ (see Fig.~\ref{fig:Disp_hex}), 
the eigenvalue with largest modulus in a given sector with
$n$ fermions is obtained by a product over the $n$ nearest to $0$ wave
vectors. Let us denote this eigenvalue by $\Lambda_n$. Then,
$\Lambda_{max}=\textrm{Max }\left\{\left|\Lambda_n\right|,\,1\leq n\leq
L\right\}$. The  eigenvalue with largest modulus is real and has approximately
all allowed $k$ lying in the interval $[-2\pi/3,2\pi/3]$. The dominant sector
has therefore $n\simeq 2L/3$ fermions. It is easy to understand the fact that the ``dominant'' fermion density is $2/3$ : it corresponds to flat height 
configurations (see Fig.\ref{fig:Hexagone}).\\
If we denote by $\Omega$ the set of wave-vectors that gives the largest
eigenvalue then the dominant eigenvector is :
\begin{equation}
 |g\ra=\left(\prod_{k \in \Omega}c_k^\dag\right)|0\ra.
\end{equation}
Let us show what is $\Omega$ in the simple case where $L=6p$. We have to
distinguish between the even and odd sectors :
\begin{equation}
\begin{array}{l}
\displaystyle{\Lambda_{\rm max}^{(e)}=\textrm{Max}\left
\{\Lambda_{2n'}\right\}=\prod_{l=p}^{5p-1}\lambda\left(-\pi+\pi\frac{2l+1}{6p}
\right)}\\\displaystyle{
\Lambda_{\rm max}^{(o)}=\textrm{Max}\left
\{\Lambda_{2n'+1}\right\}=\prod_{l=p}^{5p}\lambda\left(-\pi+ \frac{\pi
l}{3p}\right)}
\end{array}.
\end{equation}
Here, $\Lambda_{\rm max}^{(e)}>\Lambda_{\rm max}^{(o)}$
because Euler-Maclaurin expansion gives 
$\log \Lambda_{max}^{(e)}-\log \Lambda_{max}^{(o)}=\frac{\pi \sqrt{3}}{24p}+o(1/p)$.
Therefore the leading
eigenvalue corresponds to $4p=\frac23 L$ fermions and 
\begin{equation}
\Omega=\left\{-\pi+\pi\frac{2l+1}{6p} \bigg| ~ p\leq l \leq 5p-1\right\}.
\end{equation}

\subsection{Probability of a given configuration}
The dominant eigenvector has $n=2L/3$ fermions. A configuration $i$ is
represented by
\begin{equation}
|i\ra= c_{\alpha_1}^\dag \ldots c_{\alpha_n}^\dag |0\ra,
\end{equation}
and will have a probability (we use Eq.~\eqref{eq:wick} and Eq.~\eqref{E:pf}) :
\begin{eqnarray}
 p_i&=&\left|\la0|c_{\alpha_n}\ldots c_{\alpha_1}c_{k_1}^\dag \ldots
c_{k_n}^\dag|0\ra\right|^2\\
&=&\left(\frac{1}{L}\right)^n \left|\det \,(e^{-i\alpha_j
k_{j'}})_{jj'}\right|^2.
\end{eqnarray}
We get a Vandermonde determinant and $p_i$ simplifies into
Eq.~\eqref{eq:proba_hexa}.

\section{Transfer matrix for the classical dimer model on the square lattice}
\label{sec:square_tm}
\subsection{Free fermions}
\begin{figure}
\begin{center}
\includegraphics[width=8cm]{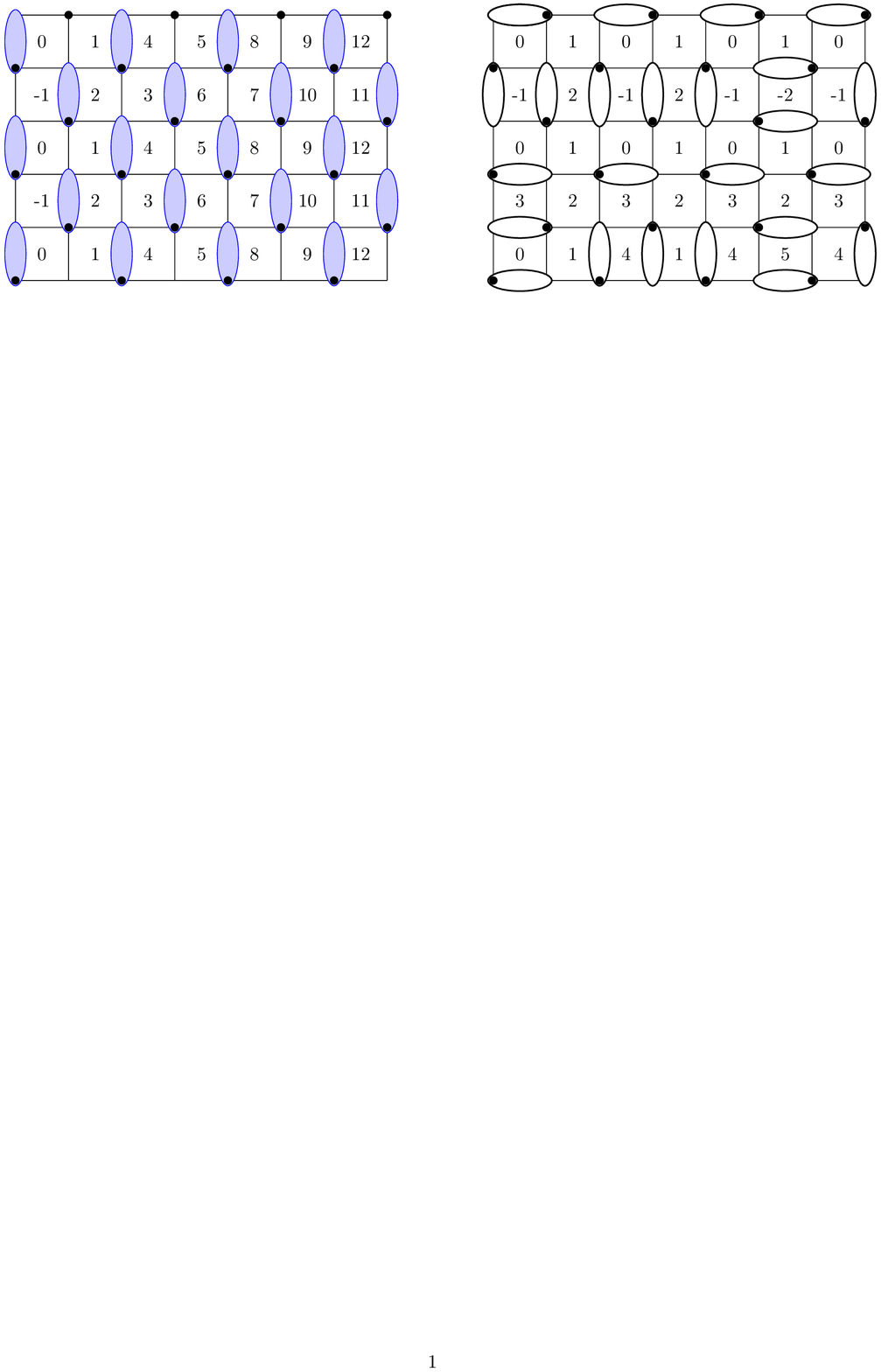}
\end{center}
\begin{center}
 \includegraphics[width=4cm]{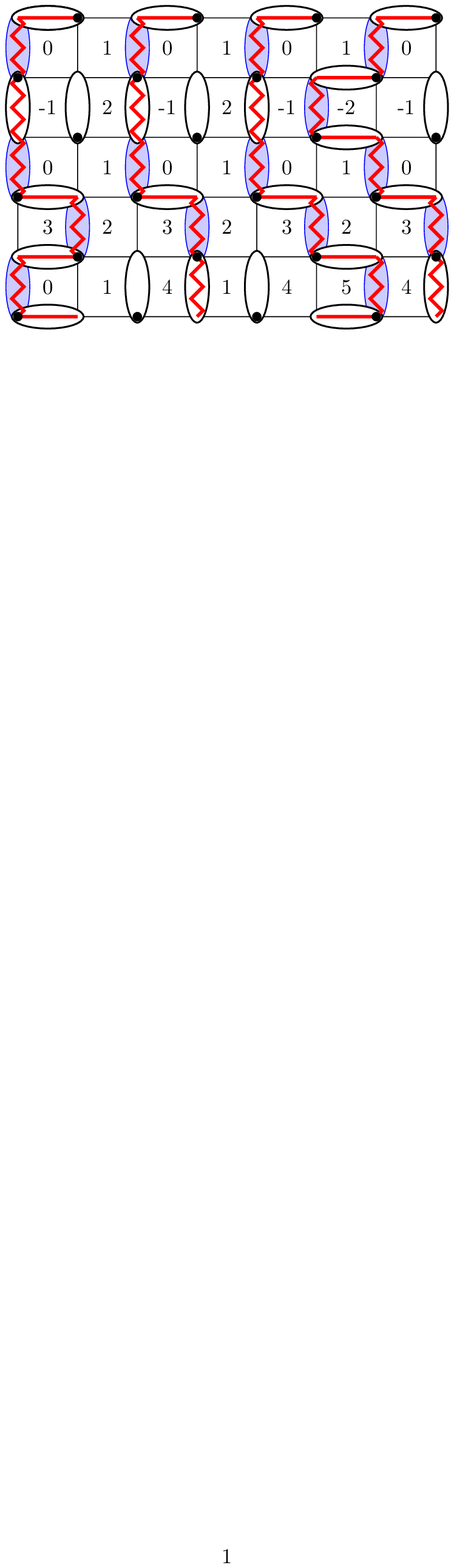}
\caption{(color online) Upper left : reference configuration. Upper right :
chosen configuration. Below : the fermions are living on the vertical edges of
the lattice and are symbolized by red zigzag lines. Edges are numbered from $0$
to $L-1$. The integers attached to each plaquette 
form a height configuration associated to the dimer covering. The rule is very similar to that of the honeycomb lattice : 
turning clockwise around the sites of the even sublattice (marked with black dots),
 the height picks a contribution equal to $+3$ when crossing a dimer, $-1$ otherwise. $h=0$ is fixed at some origin. 
For a more detailed presentation, see Ref.~\onlinecite{tmsquare}.
}\label{fig:Carre}
\end{center}
\end{figure}

We consider a square lattice with periodic boundary conditions and an even
number of columns $L$. The mapping is similar to that of the hexagonal lattice. The
reference configuration is shown in Fig.~\ref{fig:Carre}. Here, a fermion will
be defined as an even vertical edge occupied only by a reference dimer, or an
odd vertical edge occupied only by a real dimer. It can go to the left, straight
ahead, or to the right. We introduce a shift in the numbering, so that a fermion
located on site $j$ can go to $j$, $j+1$, or $j+2$.

\subsection{Diagonalization of the transfer matrix}
As for the honeycomb case, $\mathcal{T}$ is block-diagonal and invariant by
translation. It also satisfies 
\begin{eqnarray}
 \mathcal{T}|0\ra&=&|0\ra\\
\mathcal{T}c_{2j}^\dag \mathcal{T}^{-1}&=&c_{2j+2}^\dag +c_{2j+1}^\dag
+c_{2j}^\dag\\
\mathcal{T}c_{2j+1}^\dag \mathcal{T}^{-1}&=&c_{2j+2}^\dag.
\end{eqnarray}
As usual we also define Fourier space fermions :
\begin{eqnarray}
c_{0k}^\dag&=&\frac{1}{\sqrt{L/2}}\sum_{j}e^{-ik2j}c_{2j}^\dag\\
c_{1k}^\dag &=& \frac{1}{\sqrt{L/2}}\sum_j e^{-ik(2j+1)}c_{2j+1}^\dag,
\end{eqnarray}
with
\begin{eqnarray}\nonumber
k \in \bigg\{-\frac\pi2+\frac{\pi}L+ \frac{2\pi l}L\,\bigg|\;l=0,\ldots, \frac{L}2-1\bigg\}&,&\hat{n}\textrm{
even} \\\nonumber
k \in \bigg\{-\frac\pi2+\frac{2\pi l}L\,\bigg|\,l=0,\ldots, \frac{L}2-1\bigg\}&,&\hat{n}\textrm{ odd}.
\end{eqnarray}
The transfer matrix acts on them in the following way :
\begin{eqnarray}
 \mathcal{T}c_{0k}^\dag \mathcal{T}^{-1}&=&(1+e^{2ik})c_{0k}^\dag +
e^{ik}c_{1k}^\dag\\
\mathcal{T}c_{1k}^\dag \mathcal{T}^{-1}&=&e^{ik}c_{0k}^\dag.
\end{eqnarray}
To diagonalize $\mathcal{T}$, it is therefore sufficient to diagonalize a
$2\times 2$ matrix :
\begin{equation}
 M=\left(\begin{array}{cc}1+e^{2ik}&e^{ik}\\e^{ik}&0\end{array}\right).
\end{equation}
If one sets :
\begin{eqnarray}
 \tan \theta_k&=&\sqrt{1+\cos^2 k}-\cos k\\
\lambda_{\pm}(k)&=&e^{ik}\left(\cos k \pm \sqrt{1+\cos^2 k}\right)\\
b_{+k}^\dag&=& \cos \theta_k c_{0k}^\dag +\sin \theta_k c_{1k}^\dag \\
b_{-k}^\dag &=&-\sin \theta_k c_{0k}^\dag +\cos \theta_k c_{1k}^\dag,
\end{eqnarray}
then
\begin{eqnarray}
 \mathcal{T}b_{+k}^\dag \mathcal{T}^{-1}&=&\lambda_+(k) b_{+k}^\dag\\
\mathcal{T}b_{-k}^\dag \mathcal{T}^{-1}&=&\lambda_-(k) b_{-k}^\dag,
\end{eqnarray}
which gives us all the eigenvalues and eigenvectors of $\mathcal{T}$. It is also
possible to express explicitly $\mathcal{T}$ :
\begin{equation}\nonumber
 \mathcal{T}\!=\prod_k \!\left[1\!+\!(\lambda_+(k)-1)b_{+k}^\dag
b_{+k}\right]\!\!\left[\!1+\!(\lambda_-(k)-1)b_{-k}^\dag b_{-k}\right].
\end{equation}
\subsection{Largest eigenvalue and dominant eigenvector}
\begin{figure}
\begin{center}
\includegraphics[width=8.5cm]{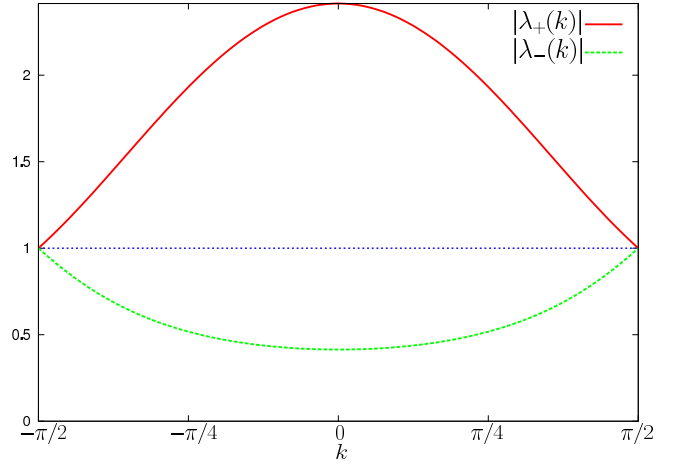}
\end{center}
\caption[1]{(color online) One-particle eigenvalues $\lambda_+$ and $\lambda_-$
as a function of $k$}\label{fig:Disp_square}
\end{figure}

We assume for simplicity that $L$ is a multiple of $4$. Noticing (see
Fig.~\ref{fig:Disp_square}) that $\forall k, |\lambda_+(k)|\geq 1$ and also
$|\lambda_-(k)|\leq 1$, we can deduce that only the $\lambda_+(k)$ will
contribute to the largest eigenvalue. In Ref.~\onlinecite{tmsquare} it is shown
that
\begin{equation}
 \lambda_{\rm max}=\prod_{k\in \Omega}\lambda_+(k),
\end{equation}
where $\Omega = \{-\frac{\pi}{2}+\frac{\pi}{L}+\frac{2\pi}{L}l\;|\; l=0\ldots
L/2-1 \}$. The leading sector has an even number of fermions $L/2$ and the
dominant eigenvector will be given by 
\begin{equation}
 |g\ra=\left(\prod_{k\in \Omega}b_{+,k}^\dag \right)|0\ra.
\end{equation}

\subsection{Probability of a given configuration}
The dominant eigenvector has $n=L/2$ fermions. A configuration $i$ is
represented by:
\begin{equation}
 |i\ra=c_{\alpha_1}^\dag \ldots c_{\alpha_n}^\dag |0\ra,
\end{equation}
and will have a probability (using Eq.~\eqref{eq:wick} and also Eq.~\eqref{E:pf}):
\begin{eqnarray}
 p_i&=&\left|\la0|c_{\alpha_n}\ldots c_{\alpha_1}b_{+,k_1}^\dag \ldots
b_{+,k_n}^\dag|0\ra\right|^2\\
&=&\left(\frac{2}{L}\right)^n \left|\det m_{jj'}\right|^2\label{eq:square_det}
\end{eqnarray}
where :
\begin{equation}
 m_{jj'}=\left\{\begin{array}{lll}\cos \theta_{k_{j'}} e^{i\alpha_j
k_{j'}}&,&\alpha_j \textrm{ even}\\\sin \theta_{k_{j'}} e^{i\alpha_j
k_{j'}}&,&\alpha_j \textrm{ odd}\end{array}\right..
\end{equation}
This determinant is slightly more complicated than on the honeycomb lattice, and
cannot be further simplified.

\section{2D Coulomb gas on a circle}
\label{sec:gaudin_annexe}
We consider a Gaudin model with $n$ charges dispatched on a circle with $L$
sites.
\subsection{Scaling of the ground state}
We study the special case where $L/n \in \mathbb{N}$ and we set $\rho=n/L$. We
denote by $p_0(\beta)$ the probability associated with the ground-state
configuration. It corresponds to the case where the distance between each
charges is maximal. Therefore it is obtained when all charges lie on the edges
of a polygon. 
Hence, 
\begin{eqnarray}\nonumber
p_0(\beta)&=&\frac{1}{Z_n^{(L)}(\beta)}L^{-n\beta/2}\prod_{1\leq k<l\leq n}
|e^{2il\pi/n}-e^{2ik\pi/n}|^\beta.
\end{eqnarray}
Using the formula $\prod_{l=1}^{n-1}\left(1-e^{2il\pi/n}\right)=n$, we get :
\begin{eqnarray}
 p_0(\beta)&=&\frac{\rho^{n\beta/2}}{Z_n^{(N)}(\beta)}. 
\end{eqnarray}
In the special case where $\beta$ is an even integer and $\beta<\frac{2L}{n-1}$,
it is possible to use Eq.~\eqref{01-gaudin} and the subleading term of $-\log
p_0(\beta)$ can easily be found :
\begin{equation}
 -\log p_0(\beta)=an+\frac{1}{2}\log \frac{\beta}{2}.
\end{equation}
This result for the universal part of the probability, $\gamma=\frac{1}{2}\log
\frac{\beta}{2}=\log R$, does not depend on $L$ nor $n$ (no finite-size
effects).

\subsection{Existence of a gap in the thermodynamical limit}
The first excitation is obtained from the ground-state configuration by moving
one particle to the next site, while keeping all the others in place. The
associated probability $p_1(\beta)$ will be given by :
\begin{eqnarray}\label{eq:gap}
 \frac{p_1(\beta)}{p_0(\beta)}=\left[\prod_{l=1}^{n-1}\frac{\sin
\left(\frac{l\pi}{n}-\frac{\rho
l\pi}{n}\right)}{\sin\left(\frac{l\pi}{n}\right)}\right]^\beta.
\end{eqnarray}
In the limit $n\rightarrow +\infty$ it is possible to expand the $\sin$ :
\begin{eqnarray}
 \frac{p_1(\beta)}{p_0(\beta)}=\left[\prod_{l=1}^{n-1}\left(1-\frac{\rho\pi}{n}
\cot (l\pi/n)\right)\right]^\beta.
\end{eqnarray}
We then consider $P_n(x)=\prod_{l=1}^{n-1}\left[1-x\cot (l\pi/n)\right]$. The
trick to calculate $P_n(x)$ is to introduce another polynomial $Q_n(x)$ which
satisfies :
\begin{equation}
 Q_n(\tan t)=\frac{\sin nt}{(\sin t) \cos^{n-1}t}.
\label{eq:trick}
\end{equation}
$Q_n(x)$ and $P_n(x)$ are of the same degree, and share the same zeros : they
have to be proportional. $P_n(0)=1$ and $Q_n(0)=n$ yields
$P_n(x)=\frac{1}{n}Q_n(x)$. Using Eq.~\eqref{eq:trick} and Moivre's formula, we
get :
\begin{equation}
 P_n(x)=\frac{1}{n}\sum_{k=0}^{[(n-1)/2]}(-1)^k C_{n}^{2k+1}x^{2k}.
\end{equation}
Therefore $P_n(\frac{\pi\rho}{n})$ reduces in the limit $n\rightarrow \infty$
to:
\begin{eqnarray}
P_n(\frac{\pi\rho}{n})=\frac{1}{\pi\rho}\sin \left(\pi\rho\right).
\end{eqnarray}
Finally, 
\begin{equation}
 \frac{p_1(\beta)}{p_0(\beta)}=\left[\frac{1}{\pi\rho} \sin
(\pi\rho)\right]^\beta.
\end{equation}
So that there is a finite gap in the thermodynamical limit:
\begin{equation}
 \Delta E=E_1-E_0=-\log \left[\frac{1}{\pi\rho}\sin (\pi\rho)\right].
\end{equation}
This calculation can easily be extended (in the thermodynamical limit) to any
configuration deduced from the ground state by moving a finite number of
particle. For the corresponding RK wave-function, $\Delta E$ gives an
information about the first gap of the reduced density matrix (entanglement
spectrum).

\section{Ground state of the Ising chain in a transverse field}
\label{sec:ictf}
\subsection{Diagonalization}
\label{sec:ictf_diag}
We consider the Hamiltonian of an Ising chain in a transverse field with an
\textit{even} number of sites $L$.
\begin{equation}
 \mathcal{H}=-\mu \sum_{j=0}^{L-1}\sigma_j^x
\sigma_{j+1}^x-\sum_{j=0}^{L-1}\sigma_j^z.
\label{eq:ising_hamiltonian}
\end{equation}
Using a Jordan-Wigner transformation
\begin{eqnarray}
 \sigma_j^+=\frac{\sigma_j^x+i\sigma_j^y}{2}&=&c_j^\dag \exp\left(i\pi
\sum_{l=0}^{j-1}c_l^\dag c_l\right)\\
\sigma_j^z&=&2c_j^\dag c_j-1,
\end{eqnarray}
$\mathcal{H}$ is rewritten as
\begin{eqnarray}\nonumber
 \mathcal{H}=&-&\sum_{j=0}^{L-1}(2c_j^\dag c_j-1)-\mu \sum_{j=0}^{L-2}(c_j^\dag
-c_j)(c_{j+1}^\dag+c_{j+1})\\
&+&\mu (c_{L-1}^\dag -c_{L-1})(c_0^\dag+c_0)e^{i\pi \mathcal{N}},
\label{eq:Ising_fermions}
\end{eqnarray}
where $\mathcal{N}=\sum_{j=0}^{L-1}c_j^\dag c_j$ is the fermion number operator.
Since $\mathcal{P}=\prod_{j=0}^{L-1}\sigma_j^z=\pm 1$ is a conserved quantity,
$\mathcal{H}$ may be separately diagonalized in two  sectors. Here we are only
interested in the ground-state of the chain. In the basis of the eigenstates of
the $\sigma_j^z$, all off-diagonal elements are negative, and it lies in the
sector $\mathcal{P}=+1$ (Perron-Frobenius theorem). Using $\mathcal{P}=\exp
(i\pi \mathcal{N})$ and the last term of Eq.~\eqref{eq:Ising_fermions}, we see
that in this sector, one has to keep configurations with an even number of
fermions satisfying anti-periodic boundary conditions
\begin{equation}
 c_L^\dag=-c_0^\dag.
\end{equation}
To take advantage of the translational invariance, we introduce Fourier-space
fermions
\begin{equation}
 c_k^\dag = \frac{1}{\sqrt{L}}\sum_{j=0}^{L-1}e^{-ikj} c_j^\dag
\end{equation}
where $k \in \{(2l+1)\frac{\pi}{L}\;|\;-L/2\leq l \leq L/2-1\}$ are the $L$
wave-vectors. The Hamiltonian becomes
\begin{eqnarray}\nonumber
\mathcal{H}=&L&-\sum_k 2(1+\mu \cos k)c_k^\dag c_k\\
 &+&\mu \sum_k \left(i\sin k c_k^\dag c_{-k}^\dag -i\sin k c_{-k}c_k\right).
\label{eq:Ising_fourier}
\end{eqnarray}
This expression can be diagonalized using a Bogoliubov transformation
\begin{eqnarray}\label{eq:bogo}
 c_k^\dag&=& \cos \theta_k b_k -i\sin \theta_k b_{-k}^\dag,\\
\theta_{-k}&=&-\theta_k,
\end{eqnarray}
provided the following condition is satisfied :
\begin{equation}
 \tan 2\theta_k=\frac{\mu \sin k}{1+\mu \cos k}.
\label{eq:diag_condition}
\end{equation}
We then obtain :
\begin{eqnarray}
 \mathcal{H}&=&\sum_k \varepsilon(k) \left[b_k^\dag b_k-1/2\right]\\
\frac{\varepsilon(k)}{2}&=&(1+\mu \cos k)\cos 2\theta_k +\mu \sin k \sin
2\theta_k.
\end{eqnarray}
In the following, we want the vacuum of the $b_k$ to be the ground-state of
$\mathcal{H}$, which is true only if $\varepsilon(k)>0 \; \forall k$. One also
has to take into account the indetermination of $\theta$ modulo $\pi$. There are
two cases :
\begin{itemize}
 \item $\mu \leq 1$ : Here $1+\mu \cos k$ is always positive. We choose
\begin{equation}\label{eq:theta1}
 \theta_k=\frac{1}{2}\arctan \left(\frac{\mu \sin k}{1+\mu \cos k}\right),
\end{equation}
and the energy spectrum is given by :
\begin{equation}
 \varepsilon(k)=2 \sqrt{1+2\mu \cos k +\mu^2}.
\end{equation}
 \item $\mu>1$ : Here, one has to be careful because $1+\mu \cos k$ can vanish
and change sign. A generic solution of Eq.~\eqref{eq:diag_condition} is 
\begin{equation}\label{eq:theta2}
 \theta_k=\frac{1}{2}\arctan \left(\frac{\mu \sin k}{1+\mu \cos
k}\right)+\frac{\pi}{2}q_k \quad,
q_k \in \mathbb{Z}.
\end{equation}
The eigenenergies are given by :
\begin{equation}
 \varepsilon(k)=(-1)^{q_k}\textrm{sgn}(1+\mu\cos k)2\sqrt{1+2\mu \cos k +\mu^2}.
\end{equation}
$1+\mu \cos k $ changes sign at $k=\pm k_c=\pm \arccos(-1/\mu)$. A possible
choice is therefore :
\begin{equation}
 q_k= \left\{\begin{array}{ccc}-1&,& k<-k_c\\
		0&,&-k_c\leq k\leq k_c\\
              1&,& k>k_c
		
             \end{array}\right..
\end{equation}
\end{itemize}
\subsection{Probability of a given configuration}
\label{se:ictf_proba}
Since $|0\ra$ is the ground state of the chain, the probability of each
configuration $i$ is (in the $\sigma^z$ basis) 
\begin{equation}
 p_i=\big\la 0\big|P_1^{\uparrow/\downarrow}P_2^{\uparrow/\downarrow}\ldots
P_L^{\uparrow/\downarrow}\big|0\big\ra,
\end{equation}
where $P_j^\uparrow$ (resp. $P_j^\downarrow$) is the projector onto the
$|\uparrow\rangle_j^z$ (resp. $|\downarrow\rangle_j^z$) state :
\begin{eqnarray}
P^{\uparrow}_j=c_j^\dag c_j && P^{\downarrow}_j=c_jc_j^\dag.
\end{eqnarray}
Using Wick's theorem, $p_i$ reduces to a Pfaffian. To compute it, we need to
calculate four types of contractions : $\la c_j^\dag c_{j'}\ra$, $\la
c_jc_{j'}^\dag\ra$, $\la c_j^\dag c_{j'}^\dag\ra$ and $\la c_jc_{j'}\ra$. This
can be done by expressing back the Jordan-Wigner fermions in terms of the
Bogoliubov fermions.
\begin{eqnarray}
 \la c_j^\dag c_{j'}\ra&=&\frac{1}{L}\sum_k \cos^2 \theta_k \cos
\left[k(j-j')\right]\\
\la c_j c_{j'}^\dag\ra&=&\frac{1}{L}\sum_k \sin^2 \theta_k \cos
\left[k(j-j')\right]\\
\la c_j^\dag c_{j'}^\dag\ra&=&\frac{1}{L}\sum_k \sin \theta_k \cos \theta_k
\sin\left[k(j'-j)\right]\\
\la c_j c_{j'}\ra&=&\frac{1}{L}\sum_k \sin \theta_k \cos \theta_k
\sin\left[k(j-j')\right].
\end{eqnarray}
If we write a generic projector as $P_j^{\uparrow/\downarrow}=a_{2j-1}a_{2j}$,
where $a$ is either $c$ or $c^\dag$, then
\begin{equation}
 p_i=\textrm{Pf }\left(\la a_j a_{j'}\ra\right)_{1\leq i,j\leq 2L}.
\end{equation}
Notice that it is also possible to compute $p_i$ when $L$ is odd. The only
difference is that the fermion number operator satisfies
\begin{equation}
 \exp(i\pi \mathcal{N})=-\mathcal{P}.
\end{equation}
Therefore, one has to take periodic boundary conditions $c_L^\dag=c_0^\dag$, and
only keep configurations with an odd number of fermions. The wave-vectors are
now in the set $\{2l\pi/L\;|\;-L/2\leq l \leq L/2-1\}$. Since the dispersion
relation $\varepsilon(k)$ is minimum for $k=-\pi$, the ground-state wave function
is 
\begin{equation}
 \left|\tilde{0}\right\ra=b_{-\pi}^\dag |0\rangle,
\end{equation}

and the probabilities
\begin{equation}
  p_i=\big\la
\tilde{0}\big|P_1^{\uparrow/\downarrow}P_2^{\uparrow/\downarrow}\ldots
P_L^{\uparrow/\downarrow}\big|\tilde{0}\big\ra
\end{equation}
will also be given by Pfaffians.
\subsection{Configuration with highest probability}
\label{sec:ictf_pmax}
We study the case where $L$ is even. The configuration with highest probability is attained by the ferromagnetic configuration 
\begin{equation}
 |i_0\ra=|\uparrow \uparrow \uparrow \ldots\ra_z.
\end{equation}
Defining 
\begin{equation}
 P=\prod_{j=0}^{L-1}c_j^\dag c_j,
\end{equation}
$p_0$ is expressed as
\begin{equation}
 p_0=\la 0|P |0\ra.
\end{equation}
$P$ is a projector onto a state with $L$ fermions.
\begin{equation}
 Q=\prod_{k}c_k^\dag c_k
\end{equation}
 is also a projector onto a state with $L$ fermions. Since
 there is only \textit{one} state of the Hilbert space with $L$ fermions, $P$ and $Q$
 are in fact identical. Using Eq.~\eqref{eq:bogo}, we therefore get :
\begin{equation}
 p_0=\prod_k \cos \theta_k,
\end{equation}
where $\theta_k$ is given by Eq.~\eqref{eq:theta1} or Eq.~\eqref{eq:theta2}. At the critical point ($\mu=1$), $\theta_k=k/4$.


\end{document}